%% file: paper.tex
\documentclass{imsart}

\usepackage{amsthm}
\usepackage{amsmath}
\usepackage{multirow}
\usepackage{graphicx}
\usepackage{subfigure}
\usepackage{natbib}
\usepackage{tikz}
\usetikzlibrary{shapes.geometric, arrows, fit}
\usepackage{flowchart}
 \tikzset{  bigbox/.style={draw, inner sep=20pt,  label={[shift={(-1ex,-6ex)}]   #1} }       }

 \startlocaldefs
\newtheorem{thm}{Theorem}
\newtheorem{fact}{Fact}
\newcommand{\ds}{\displaystyle}
\newcommand{\nn}{\nonumber}
\newcommand{\xo}{ \frac{D_j}{\lambda}+\frac{E_j^2}{\lambda^2}  }
\newcommand{\squareo}{ \beta \langle h_1, s \rangle + \langle r(g), r(s) \rangle }
\newcommand{\rooto}{ \sqrt{\beta^2 + \| r(g) \|^2} }
\newcommand{\goes}{\rightarrow}
\newcommand{\inn}[2]{\langle {#1}, {#2} \rangle}

\newcommand{\G}{\mathcal{G}}
\endlocaldefs
 
\begin{document}
\begin{frontmatter}
\title{A clustering method for misaligned curves}
\runtitle{Curve clustering}

\author{\fnms{Yu-Hsiang} \snm{Cheng}\ead[label=e1]{96354501@nccu.edu.tw}}
\address{\printead{e1}}
\affiliation{Academia Sinica, Taiwan R.O.C.}
 \and
 \author{\fnms{Tzee-Ming} \snm{Huang}\corref{}\ead[label=e2]{tmhuang@nccu.edu.tw}\thanksref{t1}}
\thankstext{t1}{Corresponding author.} 
\address{\printead{e2}}
\affiliation{National Chengchi University, Taiwan R.O.C.}
 \and
 \author{\fnms{Su-Fen} \snm{Yang}\corref{}\ead[label=e3]{yang@mail2.nccu.tw}}
\address{\printead{e3}}
\affiliation{National Chengchi University, Taiwan R.O.C.}
\runauthor{Cheng and Huang and Yang}

\end{frontmatter}

\begin{abstract}
We consider the problem of clustering misaligned curves.  According to our similarity measure, two curves are considered similar if they have the same shape after being aligned, and the warping function does not differ from the identity function very much.
 A clustering method is proposed, which updates curves so that similar curves become more similar, and then combines curves that are similar enough to form clusters. The proposed method needs to be used together with a clustering index and a set of combination thresholds.
 Simulation results are presented to demonstrate the performance of this approach under different parameter settings and clustering indexes. Two real data applications are  included. 
\end{abstract}

 \section{Introduction} \label{sec:introduction}
\input{introd_align}

\section{Methodology} \label{sec:method}
\input{method_new}

\section{Simulation studies} \label{sec:simu}
 \input{simulation}

\section{Data applications}  \label{sec:data}
 \input{data_analysis}
\section{Discussion and suggestions} \label{sec:dis}
 \input{conclusion}

\section{Proofs} \label{sec:proofs}
 \input{proof}


\input{refe.bbl}


\end{document}

%% file: introd_align.tex
Functional data are often observed over time and
it is usual that a set of data curves show a common pattern with some variation in time.
Before performing further analyses on data curves, such as estimating the common pattern, synchronizing the observed curves is necessary. Thus, curve alignment is an important problem in functional data analysis.

In the literature, many curve alignment methods have been proposed. One approach for curve alignment is landmark registration in \cite{Knei:1992}. Landmarks are selected characteristics of curves, such as peaks or valleys. Another approach for curve alignment is continuous monotone registration, in which smooth monotone time transformations or warping functions are used to align individual curves to target curves. Curve alignment methods based on continuous monotone registration can be found in \cite{Silv:1995}, \cite{Rams:1998}, \cite{Knei:2000}, \cite{Gerv:2004}, and \cite{Tele:2008}. James \cite{Jame:2007} proposed a curve alignment method based on moments, which is a hybrid of the landmark approach and the continuous monotone registration approach.

In the past few years, some authors have  investigated the clustering problem for misaligned curves. \cite{Tang:2009} proposed a two-step clustering method. In the first step, curves are aligned using estimated cluster-specific warping functions. In the second step, the aligned curves can be clustered using  any existing clustering method, such as  $k$-means clustering or hierarchical clustering.  In this method, it is assumed that the warping functions are non-linear and satisfy the boundary condition.   \cite{Liu:2009} proposed the SACK model and  provided an estimation procedure using the EM algorithm. Later,  \cite{Sang:2010} proposed a $k$-means algorithm for clustering misaligned curves.  In contrast to the approach in \cite{Tang:2009}, in both \cite{Liu:2009} and \cite{Sang:2010}, linear warping functions are considered, and curve alignment and clustering are performed simultaneously, as summarized in Table \ref{table:1}.

\small
\begin{table*}[h] \renewcommand{\arraystretch}{1} \small
\caption{Comparison between our method and other clustering methods}
\label{table:1}
\begin{center}
    \begin{tabular}{ccccc}
    \hline  \hline
     &   k-means  & SACK   &  Two-step clustering   &  Our method \\
    \hline
    Simultaneous clustering    &    &   &   &  \\
       and alignment  &  Yes & Yes & No & Yes\\
 &    &   &   &  \\
    Linear warping function & Yes & Yes & No & No \\
   \hline \hline
\end{tabular}
\end{center}
\end{table*}
\normalsize
 
 In our option, it is better to perform alignment and clustering simultaneously. For instance,  when we use the two-step clustering in \cite{Tang:2009}, not all curves in the same cluster can be aligned well, as shown in Figure \ref{tangplot}.  It seems more efficient to perform alignment and clustering simultaneously.  For using linear warping functions or  nonlinear warping functions, we do not prefer one to the other and a suitable choice should be made based on  the nature of  data. 

In this paper, we provide a method for clustering misaligned curves under the same assumptions for warping functions as in \cite{Tang:2009}. For the proposed method, we include a parameter $\lambda_0$ to adjust the  penalty for large time variation. If it is desirable to put curves with different degrees of time variation into different groups, this can be done  using a large $\lambda_0$.  We organize this paper as follows. In Section 2, the details of this proposed method and theoretical result are given. Some results of simulation studies and analyses for two real data sets are presented in Sections 3 and 4. Discussion and suggestions are given in Section 5. Proofs are given in Section 6.

%% file: method_new.tex
In this section, we will first describe the set-up of the clustering problem and introduce the similarity measure in Section \ref{section:m1}. Next, we give an overview of the proposed clustering process  in Section \ref{section:overview}.  A theoretical result for curve updating is given in Section \ref{section:theorem}.
\subsection{Similarity measure} \label{section:m1}
We consider the problem of clustering $m_0$ data curves $y_1$, $\ldots$, $y_{m_0}$, where the curves are observed at time points $0 = t_1 < t_2 \cdots < t_n = 1$. For $i = 1$, $\ldots$,  $m_0$ and $j =1$, $\ldots$, $n$, let $y_i(t_j)$ denote the observed value of curve $y_i$ at time point $t_j$. We assume that these curves can be modelled as
\[ 
 y_i(t_j) = f_{0,i}(t_j) + \varepsilon_{ij},
\]
where $f_{0,i}$ is called the shape function of curve $y_i$ and $\varepsilon _{ij}$s are independent errors with mean zero. The goal of clustering is to assign curves into groups so that similar curves are in the same group, and the curve similarity measure will be introduced in this section. 

For two curves with shape functions $f_0$ and $g_0$ that do not need to be aligned, a usual similarity measure is 
\[
 r(f_0,g_0) = \frac{\langle f_0, g_0 \rangle}{ \| f_0 \|  \|g_0\| },
\]
where 
\[
\langle f_0, g_0 \rangle = \int_0^1 (f_0(x) -Ef_0)( g_0(x)- Eg_0) dx,
\]
$Ef_0=\int_0^1f_0(x)dx$, $Eg_0 = \int_0^1 g_0(x)dx$, $\| f_0 \|= \sqrt{ \langle f_0, f_0 \rangle}$ and $\| g_0 \|= \sqrt{ \langle g_0, g_0 \rangle}$.
 Note that $r(f_0, g_0) \leq 1$ and $r(f_0, g_0)=1$  means that the two curves with shape functions $f_0$ and $g_0$ have the same shape (up to a scale and level change).  

Our curve similarity measure is based on the similarity measure $r$ for the warped curves, and we consider warping functions in  the space
\[ \mathcal{M} = \{ \psi: \psi \in \mathcal{M}_0: \psi(0)=0, \psi(1)=1 \mbox{ and } \psi^{-1} \in \mathcal{M}_0 \}, \]
where $\mathcal{M}_0$ is the space of continuously differentiable increasing functions defined on $[0,1]$. The boundary constraint $\psi(0)=0$ and $\psi(1)=1$ can also be found in \cite{Rams:1997} 
and is called the common endpoints condition in \cite{Tang:2009}.  Note that warping functions satisfying the boundary constraint cannot be linear unless they are equal to the identity function, and as such, we consider nonlinear warping functions.

Below we will define our curve similarity measure. For two curves with shape functions $f$ and $g$, and for  $\psi$ in $\mathcal{M}$, define
\begin{equation}\label{eq:similar1}
\rho^*(f, g| \psi) = r(f, g \circ \psi) - \lambda_0 \bigg(\int_0^1 \Big(\frac{d}{dt}\psi(t)-1\Big)^2 dt\bigg)
\end{equation}
and $\rho(f, g| \psi) = \Big(\rho^*(f, g| \psi) + \rho^*(g, f| \psi^{-1})\Big)/2$, where $\lambda_0$ is a non-negative parameter.
Then,   the similarity measure between $f$ and $g$ is defined as
\begin{equation*} \label{eq:similar}
\rho(f,g)=\max_{\psi \in \mathcal{M}} \rho(f, g|\psi).
\end{equation*}
Let $ \psi_0=\mbox{arg} \max_{\psi \in \mathcal{M}} \rho(f, g|\psi)$, then we use $\psi_0$ as the warping function when aligning $f$ to $g$ and use $\psi_0^{-1}$  as the warping function when aligning $g$ to $f$.

Note that our similarity measure depends the parameter $\lambda_0$ in (\ref{eq:similar1}). $\lambda_0$ controls the degree of time variation. For a warping function $\psi \in \mathcal{M}$, the integral $\displaystyle \int_0^1 \Big(\frac{d}{dt}\psi(t)-1\Big)^2 dt =0$ implies that $\psi$ is the identity function. Using a large $\lambda_0$ thus gives a large penalty for using warping functions that deviate from the identify function, and thus when the curves are clustered, the resulting time variation within the same group is expected to be limited. 
 A similar penalty term for the warping function can be found in Section 5.4.2 in \cite{Rams:1997}.  Some authors argue that in curve clustering, curve alignment is not always necessary  if one would like to consider time variation as a clustering factor (\cite{Jacq:2014}).  
In that case, one can use a large $\lambda_0$ in our similarity measure.

For convenience in evaluating the similarity measure, all data curves and warping functions are approximated using splines. We treat the approximate data curves as curves without errors and will not distinguish between the shape function of a data curve and the data curve itself hereafter.

\subsection{Clustering method} \label{section:overview}
 In our clustering method, curves are updated so that curves that are similar enough become more similar and then eventually can be combined to form clusters. 
In the problem of clustering points, the idea of updating points can be found in  \cite{Fuku:1975}, \cite{Chen:2007}, and \cite{Shiu:2012}. Chen and Shiu (2007, 2012) proposed a self-updating algorithm where points are moved toward their neighbors to form clusters automatically. Our approach is similar to Chen and Shiu's approach since
in both our method and Chen and Shiu's algorithm, curves (or points) are updated using weighted averages. However, the weighting schemes are different.  Our weighting scheme is based on Theorem \ref{thm1} in Section \ref{section:theorem}.

\begin{figure*}[h]
\begin{center}
\begin{minipage}[b]{14.5cm}
\begin{tikzpicture}[ font={\sf \small}]
\def \smbwd{2cm}
\node(S3A1) at (-2.2,1.2) [draw, terminal,minimum width=\smbwd, minimum height=1cm] {(A) Curve combination};  
\node(S3A2) at (-2.2,-1.8) [draw, terminal,minimum width=\smbwd, minimum height=1cm] {(B) Curve updating};  
\node(S3A3) at (-0.06,-4.4) [draw, terminal,minimum width=\smbwd, minimum height=1cm] {(D) Computing clustering indexes for all candidate results from (C) to obtain the final clustering result};  
\node(S3A2star) at (1.4,-0.5) [draw, terminal,minimum width=\smbwd, minimum height=1cm] {(C) Obtaining a candidate clustering result};  
\coordinate (point1) at (-3.1,-1.3);
\coordinate (point11) at (-3.1,0.7);
 
 \coordinate (point3) at (-5.2,2.4);
\coordinate (point2) at (5.23,-1.8);

 \coordinate (point31) at (-7.2,-3.4);
\coordinate (point21) at (7.23,-4.4);

 \node[bigbox={Tasks in each iteration $\mbox{:}$ (A)(B)(C) }, fit=(point2)(point3),] (D) {};
  \node[bigbox={Task after the iterative process is complete $\mbox{:}$ (D)}, fit=(point21)(point31),] (E){};
 
\draw[->] (S3A1) -- (S3A2);
\draw[->] (S3A1) -- (S3A2star);
\draw[->] (point1)--(point11);
\end{tikzpicture}
\end{minipage}
\end{center}
\caption{Diagram of the clustering procedure for a given combination threshold $c^*$} \label{fig:process}
\end{figure*}
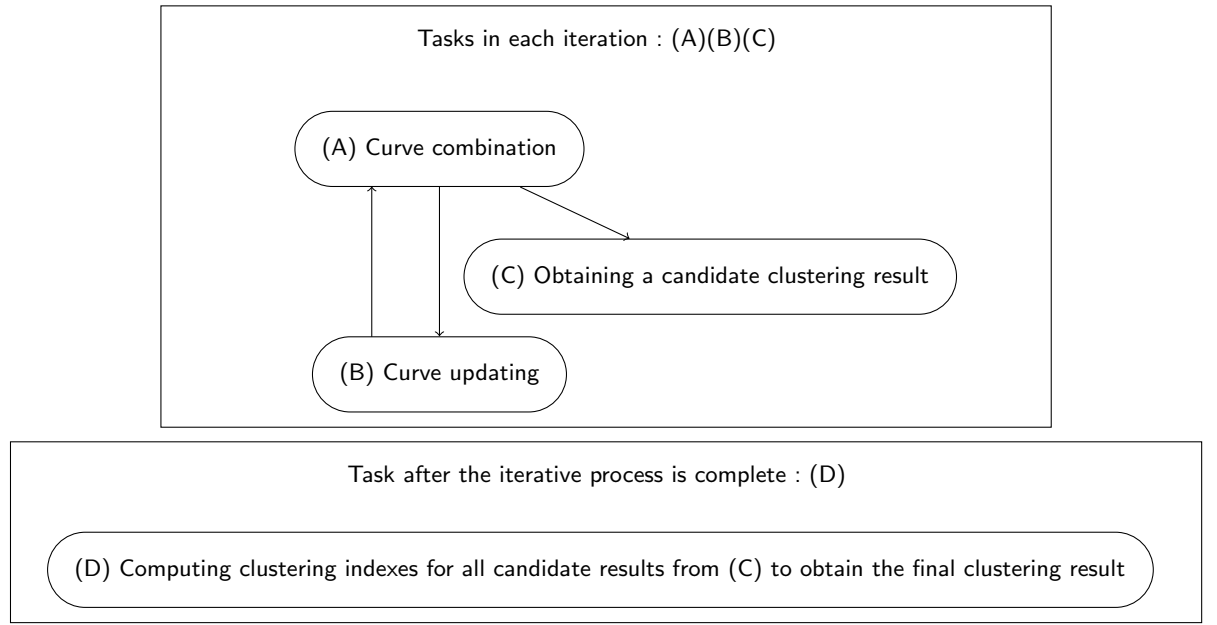

To implement our clustering method, we need to choose a set of curve combination thresholds $S_c$ and a clustering index such as the Silhouette coefficient  in \cite{Rous:1987}.  When $S_c$ contains several threshold values, we obtain the clustering result for each threshold value and then determine the final clustering result based on the clustering index. For most of our simulation studies, $S_c$ consists of 4 points near the 75\% quantile of similarity measures of original curves excluding one's.
Below we describe our clustering method when $S_c$ has only one threshold value $c^*$.   First, we start an iterative process, where in each iteration, we perform
\begin{itemize}
\item[(A)] curve combination and
\item[(B)] curve updating.
\end{itemize}
For curve combination, two curves can be combined if their similarity measure exceeds the given threshold $c^*$. After curves are combined, curves will be updated. The iterative process stops when the average curve similarity measure remains stable. Note that at the end of this iterative process, it is possible that all curves are combined into one curve. To obtain a final clustering result from the whole iterative process, in each iteration, if some curves are combined in Step (A), then we obtain a candidate clustering result based on the updated curves after Step (A) and before Step (B) in that iteration. Thus at the end of the  iterative process, several candidate clustering results are obtained, and the candidate result with the best clustering index is chosen as the final clustering result, where the clustering indexes are calculated based on the similarity measures of the original curves. The clustering procedure for a given combination threshold $c^*$ is given in Figure \ref{fig:process}.   Details for  (B) curve updating, (A) curve combination and (C) obtaining a candidate clustering result in each iteration are given in Sections \ref{section:updating}, \ref{sec:curve_combination} and \ref{sec:candidate} respectively.

\subsubsection{Curve updating} \label{section:updating}
In the curve updating step,  suppose that we have several curves to be updated. Then curves are updated one at a time, and before each curve is updated, all curves are normalized so that their $L^2$ norms are equal to one. Let $f_1$ denote the curve to be updated and $f_2$, $\ldots$, $f_k$ denote the rest curves.   
Then we update $f_1$ to 
\[
 f_1^* = \left( \frac{\lambda}{\lambda+1} \right) f_1 + \left( \frac{1}{\lambda+1} \right) \sum_{j=2}^k \theta_j \frac{f_j \circ \psi_j}{\|f_j \circ \psi_j\|},
\]
where $\lambda >0$, $\theta_j \in [0, 1]$ for $j \geq 2$, $\sum_{j=2}^k \theta_j = 1$, and for $2 \leq j \leq k$, $\psi_j$ is the warping function in $\mathcal{M}$ when aligning $f_1$ to $f_j$.  

To describe $\theta_j$s, we introduce some notations.
For $2 \leq j \leq k$,  define  $E_j f = \int_0^1 f(t) \frac{d}{dt}\psi_j(t)dt$ for a real value function $f$ on $[0,1]$. 
For two real valued  functions $f$ and $g$ on $[0,1]$, define
\[
  \langle f, g \rangle_j = E_j(f-E_j f)(g-E_j g)
\]
and  $\|f\|_j = \sqrt{\langle f, f \rangle_j}$.  Then for $2 \leq j \leq k$, we set  $\theta_j=0$ if 
\[
\langle f_1, f_j \circ \psi_j \rangle \leq 0,
\]
\begin{equation} \label{eq:res1}
\left\langle  \frac{f_j \circ \psi_j}{\|f_j \circ \psi_j\|} ,  \sum_{\ell=2}^k \frac{f_\ell \circ \psi_\ell - \langle f_\ell \circ \psi_\ell, f_1 \rangle f_1}{\|f_\ell \circ \psi_\ell \|} \right\rangle  \leq 0,
\end{equation}
or
\begin{equation} \label{eq:res2}
  \sum_{\ell=2}^k \left\langle  \frac{f_j \circ \psi_j}{\|f_j \circ \psi_j \|} , \frac{1}{\|f_1\|_\ell}\Big(f_\ell \circ \psi_\ell-\frac{\langle f_\ell \circ \psi_\ell, f_1 \rangle_\ell f_1}{\|f_1\|_\ell^2}\Big)\right\rangle_\ell   \leq  0.
\end{equation}
For the $\theta_j$s that are nonzero, we choose  $\theta_j$s  to be proportional to $n_j w_{1,j}^{\tau}$, where  $n_j$ is the number of original curves that are updated/combined to form the curve $f_j$, \[ w_{1,j} = \frac{\rho(f_1, f_j)}{\max \{ \rho(f_1, f_j) : \theta_j >0 \} },  \]$\tau = \log(0.5)/\log(\mbox{Ind}_{\max})$
 and $\mbox{Ind}_{\max}$ is the maximum of the similarity measures that are less than 1 for the original data curves.

$\lambda$ is computed based on the  $\theta_j$s.  To obtain $\lambda$,  let
\[
   g_0 = \sum_{j=2}^k \theta_j \frac{f_j \circ \psi_j}{\|f_j \circ \psi_j\|}, \hspace{0.5cm}
 s = \sum_{j=2}^k \frac{f_j \circ \psi_j}{\|f_j \circ \psi_j\|}, \] \[  s_0 = \sum_{j=2}^k \frac{f_j \circ \psi_j - \langle f_j \circ \psi_j, f_1 \rangle f_1}{\|f_j \circ \psi_j\|},
\]
 then $\lambda$ is the maximum of  the following two quantities $LC5$ and $LC6$:
\begin{equation} \label{eq:LC5}
  LC5 = \frac{\|g_0 - \langle g_0, f_1 \rangle f_1\|^2 \langle s, f_1 \rangle^2 - \langle g_0, s_0 \rangle^2}{2 \langle s, f_1 \rangle \langle g_0, s_0 \rangle} - \langle g_0, f_1 \rangle
\end{equation}
and
\begin{equation} \label{eq:LC6}
 LC6 =   \max \bigg(\frac{\sum_{j=2}^k \beta_j}{\sum_{j=2}^k \alpha_j}, \max_{2 \leq j \leq k} \max \Big(E_j, |B_j|\Big)\bigg),
\end{equation}
where
\[
\beta_j = \frac{1}{2}\Big(A_jE_j^2 + B_jD_j + |B_j|E_j\Big) + \frac{3}{\sqrt{2}}\Big(|A_j|+1\Big)\Big(|D_j|+E_j\Big)^2,
\]
\[
  A_j = \frac{\langle f_1, f_j \circ \psi_j \rangle_j}{\|f_1\|_j},  \hspace{0.5cm} B_j = \frac{\langle g_0, f_j \circ \psi_j \rangle_j}{\|f_1\|_j}, \hspace{0.5cm} D_j = \frac{2 \langle f_1, g_0 \rangle_j}{\|f_1\|_j^2}, \] $\ds E_j = \frac{\|g_0\|_j}{\|f_1\|_j}$,  
 and $\alpha_j = B_j - \frac{1}{2}A_jD_j$.

\subsubsection{Details for curve combination} \label{sec:curve_combination}
We  briefly state the tasks in the step of curve combination. In this step, we first put sets of similar curves into clusters, then for each cluster, align curves in the cluster to a reference curve, combine the aligned curves in the cluster to form a representative curve, and replace the curves in the cluster by the cluster representative. The representative curve is the fitted B-spline curve to the aligned data curves using weighted least square regression, where the reference curve for alignment is chosen as the curve in the cluster with the largest average of similarity measures to other curves in the cluster, and the weight for each curve in the weighted least square regression is the number of original curves  that are updated/combined to form this curve. Here two curves are considered similar if their similarity measure exceeding $c^*$.

The main difficulty in the step of curve combination is that sometimes we have a conflictive situation in clustering. For instance, suppose that we have a curve $f_1$ that is similar to two curves $f_2$ and $f_3$, but $f_2$ and $f_3$ are not similar, then it is not clear how these three curves should be combined. In such case, we make use of the clustering index to help resolve this difficulty.  Here the clustering index is computed based on the updated curves, not the original curves. For a combination result that assigns several updated curves to $p$ groups $G_1$, $\ldots$, $G_p$ for combination, let $\nu(G_1, \ldots, G_p)$ denote  the clustering index based on the updated curves. 

Below we give the steps for assigning $m$ curves  $f_1$, $\ldots$, $f_m$ into clusters for combination.

\begin{enumerate}
\item  Compute
\[
  \rho(f_i) = \sum_{j = 1, \ldots, m, \rho(f_i, f_j) >c^*}   \rho(f_i, f_j),
  \]
for $1 \leq i \leq m$.  

\item Sort the $m$ functions $f_1$, $\ldots$, $f_m$ by $\rho(f_i)$s (from largest to smallest). Let $S$ be the sequence of sorted functions.

\item Let $f^*$ be the first function in $S$. Form a new group $G_0$ including $f^*$ by carrying out the steps (a)--(c) below and then remove the curves in $G_0$ from $S$.
\begin{enumerate}
\item  Collect the functions $f_i$s in $S$ that satisfy  $\rho(f^*, f_i)$ $>c^*$, and then sort these functions by $\rho(f^*, f_i)$s (from largest to smallest). Let $P$ be the sequence of the above sorted functions. 

\item Let $G_0 = \{f^*\}$, and then add the functions in $P$ to $G_0$ in turn under the constraint that each newly added function is similar to all functions in $G_0$.

\item For every function in $G_0$, check whether it has similar function(s) outside $G_0$.
\begin{itemize}
\item If it has no similar function outside $G_0$, then $G_0$ is the new group containing $f^*$.
\item If it has similar function(s) outside $G_0$, carry out (**) to update $G_0$. The resulting $G_0$ is the new group containing $f^*$.
\end{itemize}
 \item[] (**)
For all functions in $G_0$, determine in turn whether they will stay in $G_0$ according to the following criterion: for a function $f^*_0$ in $G_0$, let $s(f^*_0)$ be the   collection of similar curves of $f^*_0$ and $s(G_0)$ be the collection of curves that are similar to all curves in $G_0$. Let $D = s(f^*_0) \cap s(G_0)^c$, where $s(G_0)^c$ is the complement of set $s(G_0)$. $f^*_0$ will stay in $G_0$ if either of  the following two conditions are satisfied:
 \begin{itemize}
 \item[(1)] $D$ is an empty set; 
 \item[(2)] $\nu(G_0 \cup \{ f^*_0 \}, \{ s^*\} ) >\nu(\{ f^*_0, s^* \}, G_0)$, where $s^*$ is the most similar curve to $f^*_0$ among  the curves in $D$.
 \end{itemize}
 \end{enumerate}

\item Repeat 3 until $S$ is empty. 
\end{enumerate}
Following the above steps, we can assign  $f_1$, $\ldots$, $f_m$ into clusters and then combine curves in the same cluster.

\subsubsection{Obtaining a candidate clustering result} \label{sec:candidate}
In this section, we give details for obtaining a candidate clustering result (Step (C) in Figure \ref{fig:process}). Note that in the curve combination step, we put some curves in clusters for combination and leave other curves outside those clusters, so we only have a partial clustering result.
For  curves that are not assigned into clusters for combination, we treat  those curves as unassigned curves and then perform further clustering to obtain a complete clustering result as a candidate clustering result.  The details are given below.

Let $\G_0$ and $S_0$ be the collections of groups and unassigned curves respectively based on the partial clustering result in curve combination. Let $p_0$ be the number of groups in $\G_0$ and $q_0$ be the number of unassigned curves in $S_0$. Also, for a  clustering result that assigns several updated curves to $p$ groups $G_1$, $\ldots$, $G_p$, let $\nu_0(G_1, \ldots, G_p)$ denote the similarity measure for the clustering result computed based on the original curves.   Then we can obtain the candidate clustering result by carrying out the following steps for  the case $p_0 \geq 2$ and $q_0 \geq 1$.
\begin{enumerate}
\item Set $\G=\G_0$ and $S=S_0$.

\item Suppose that $\G$ is the collection of $p$ groups $G_1$,  $\ldots$, $G_p$, and $S$ is nonempty. For every curve $g$ in $S$, compute $\nu_0( G_1\cup\{ g\}, \ldots, G_p)$, $\ldots$, $\nu_0(G_1, \ldots, G_p \cup \{ g\})$ and $\nu_0(G_1, \ldots, G_p, \{ g\})$. If the clustering result that includes  $G_k \cup \{ g \}$ has the largest $\nu_0$ value for some $k \in \{ 1, \ldots, p \}$, then adds $g$ to group $G_k$ and removes it from $S$.

\item Suppose that after Step 2,  $S$ remains nonempty and $S=\{ h_1, \ldots, h_{q} \}$ and $\G$ becomes the collection of $p$ groups $G_1^*$, $\ldots$, $G_p^*$. For $i = 1$, $\ldots$ $q$, compute $\nu_0(G_1^*, \ldots, G_p^*, \{  h_i \})$ and let $h^*$ be the $h_i$ with the largest $\nu_0(G_1^*, \ldots, G_p^*, \{  h_i \})$. Add the singleton $\{ h^* \}$ to $\G$ so that $\G$ includes exactly  $(p+1)$ groups: $G_1^*$, $\ldots$, $G_p^*$, $\{ h^* \}$. Remove $h^*$ from $S$.

\item Repeat the steps 2 and 3 until $S$ is empty.  

\item A complete clustering result is given by $\G$.
\end{enumerate}
The above steps for obtaining a complete clustering result $\G$ based on a partial clustering result characterized by a collection of non-singleton groups $\G_0$ and a collection of unassigned curves $S_0$ can be viewed as a function of $\G_0$ and $S_0$. We will name this function {\em cluster.1} and  denote the function output by {\em cluster.1}($\G_0$, $S_0$) based on  input $\G_0$ and $S_0$.
 This function will be also used to handle the cases other than $p_0 \geq 2$ and $q_0 \geq 1$. 

For cases other than $p_0 \geq 2$ and $q_0 \geq 1$, the details for obtaining a candidate clustering result $\G$ based on the partial clustering result from curve combination are given below.
\begin{itemize}
\item $p_0 = 0$ and $q_0 = 2$. Let  $g_1$ and $g_2$  be the two curves in $S_0$. If the two curves $g_1$ and $g_2$ are similar, then let $\G$ be the clustering result including only one group $\{ g_1, g_2\}$. Otherwise, let $\G$ be the clustering result of two singletons $\{ g_1\}$ and $\{ g_2\}$.

\item $p_0 = 0$ and $q_0 \geq 3$. Suppose that $S_0 = \{ g_i: i=1, \ldots, q_0 \}$.
\item[] (1)  For $i = 1, \ldots, q_0$,  compute  $\rho(g_i) = \sum_{j = 1}^{q_0} \rho(g_i, g_j)$. Let $g^*_1$ be the curve $g_i$ with the largest $\rho(g_i)$ and form a new group $\{ g^*_1 \}$. 

\item[] (2) Compute $\nu_0(\{ g^*_1 \}, \{  g_i\})$ for $g_i \neq g^*_1$, let $g^*_2$ be the curve $g_i$ with the largest $\nu_0(\{ g^*_1 \}, \{  g_i\})$ for $g_i\neq g^*_1$. Form the second group $\{ g^*_2 \}$. 

\item[] (3) Let $\G =$ {\em cluster.1} $\displaystyle \left( \{ \{ g^*_1 \}, \{ g^*_2 \}\},  \{ g_i, g_i\neq g^*_1, g^*_2 \} \right)$.

\item $p_0 = 1$ and $q_0 = 1$. Let $G_1$ be the only group in the collection $\G_0$ and $g_1$ be the only curve in $S_0$. Suppose that $G_1$ is composed of $m^*$ curves $f_1$, $\ldots$, $f_{m^*}$. Compute $\kappa_0 = \nu_0(G_1, \{ g_1 \})$ and $\kappa_i = \nu_0( \{ g_1, f_1 \ldots, f_{m^*} \} - \{ f_i \}, \{ f_i \})$ for $i = 1$, $\ldots$, $m^*$. If some $\kappa_i$ is larger than $\kappa_0$, put the curve $g_1$ into $G_1$ and let $\G$ be the clustering result including exactly the group $G_1 \cup \{ g_1\}$. Otherwise, let $\G$ be the clustering result including exactly $G_1$ and the singleton group $\{ g_1\}$.
\end{itemize}

\subsection{Curve updating - theoretical result}  \label{section:theorem}
The weighting scheme in the curve updating step  is based on our Theorem \ref{thm1} in this section. The theorem gives some conditions on the weights when updating a curve $f_1$ to a weighted average of $f_1$ and warped versions of other $(k-1)$ curves $f_2$, $\ldots$, $f_k$:
\[
 f_1^* = \left( \frac{\lambda}{\lambda+1} \right) f_1 + \left( \frac{1}{\lambda+1} \right) \sum_{j=2}^k \theta_j \frac{f_j \circ \psi_j}{\|f_j \circ \psi_j\|},
\]
where $\lambda >0$, $\theta_j \in [0, 1]$ for $j \geq 2$, $\sum_{j=2}^k \theta_j = 1$, and for $2 \leq j \leq k$, $\psi_j$ is the warping function in $\mathcal{M}$ when algning $f_1$ to $f_j$.   When the conditions in Theorem \ref{thm1} hold, the updated curve $f_1^*$ is more similar to other curves than $f_1$ on average in the sense that (\ref{aim}) holds.

We follow the definitions of $\langle \cdot, \cdot \rangle_j$ and $\| \cdot \|_j$ in Section \ref{section:updating} and state Theorem \ref{thm1} below.
\begin{thm} \label{thm1}
Suppose that the following conditions hold:
\begin{itemize}
\item[(C1)] $\|f_j \| = 1$ for $1 \leq j \leq k$.
\item[(C2)] $\langle f_1, f_j \circ \psi_j \rangle \geq 0$ for $2 \leq j \leq k$.
\item[(C3)] 
$\displaystyle
\left\langle \sum_{j=2}^k \theta_j \frac{f_j \circ \psi_j}{\|f_j \circ \psi_j\|} ,  \sum_{\ell=2}^k \frac{f_\ell \circ \psi_\ell - \langle f_\ell \circ \psi_\ell, f_1 \rangle f_1}{\|f_\ell \circ \psi_\ell\|} \right\rangle >0$.

\item[(C4)] \small $\displaystyle \sum_{\ell=2}^k \left\langle \sum_{j=2}^k \theta_j \frac{f_j \circ \psi_j}{\|f_j \circ \psi_j\|}, \frac{1}{\|f_1\|_\ell}\Big(f_\ell \circ \psi_\ell-\frac{\langle f_\ell \circ \psi_\ell, f_1 \rangle_\ell f_1}{\|f_1\|_\ell^2}\Big)\right\rangle_\ell$ $> 0$.

\item[(C5)] \normalsize $\lambda \geq  LC5$, where $LC5$ is given in (\ref{eq:LC5}).  
\item[(C6)] $\lambda \geq  LC6$, where $LC6$ is given in (\ref{eq:LC6}). 
\end{itemize}
Then,
\begin{equation}\label{aim}
\sum_{j=2}^k \rho(f_1^*, f_j) \geq \sum_{j=2}^k \rho(f_1, f_j).
\end{equation}
\end{thm}
The proof of Theorem \ref{thm1} is given in Section \ref{sec:proofs}.

In our curve updating step, curves are normalized so that (C1) holds, only $f_j$s satisfying (C2) will be used to update $f_1$, weights $\theta_j$s are selected so that (C3) and (C4) hold, and $\lambda$ is chosen so that (C5) and  (C6) hold.   To specify $\theta_j$s such that (C3) and (C4) hold, note that (C3) and (C4) are of the form  $\sum_{j=1}^k \theta_j C_j > 0$ for some known constants $C_j$s. We simply set $\theta_j =0$ when $C_j \leq 0$ for each $j$ to ensure that  $\sum_{j=1}^k \theta_j C_j > 0$. The requirement $C_j \leq 0$ for (C3) and (C4) corresponds to (\ref{eq:res1}) and (\ref{eq:res2}) 
respectively. There are certainly other ways for choosing $\theta_j$s  such that (C3) and (C4) hold, but we have not yet explored them.

 In  (\ref{eq:res1}) and (\ref{eq:res2}),  $f_\ell \circ \psi_\ell - \langle f_\ell \circ \psi_\ell, f_1 \rangle f_1$ and $f_\ell \circ \psi_\ell - \langle f_\ell \circ \psi_\ell, f_1 \rangle_j f_1$ are residuals of projecting the warped curve $f_\ell \circ \psi_\ell$ to the space spanned by $f_1$ with respect to the semi-inner products $\langle \cdot,  \cdot \rangle$ and $\langle \cdot,  \cdot\rangle_j$ respectively. The effect for setting $\theta_j=0$ if (\ref{eq:res1}) or (\ref{eq:res2}) holds is so that curves that are very dissimilar to the residuals on average can be excluded, so that the updated curve $f_1^*$ can be more similar to the rest of the curves on average.

%% file: simulation.tex
In this section, we present results of the simulation studies under various settings with $S_c$ (the set of combination thresholds) taken to be the set 
$\{ q_{1-a} - 0.01 + (0.01\cdot i/3)$:  $i=0$, 1, 2, 3 $\}$, 
where $q_{1-a}$ denotes the $100(1-a)$\% sample quantile of the similarity measures of original curves that are less than one.  

In Sections 3.1--3.3, we consider different types of warping functions using the Silhouette index as the clustering index and take $q_{1-a} = q_{0.75}$ for $S_c$.  Recall that for a warping function $\psi$, we assume that $\psi$ is monotone on $[0, 1]$ and satisfies the boundary condition
\begin{equation} \label{eq:warping01}
 \psi(0)=0 \mbox{ and  } \psi(1)=1.
\end{equation}
In Section 3.1, we consider warping functions satisfying the boundary condition (\ref{eq:warping01})  and the focus is on the effect of  parameter $\lambda_0$ and the unbalance class size. In Section 3.2, we consider warping functions that violate the boundary condition (\ref{eq:warping01}) slightly. In Section 3.3, we consider both linear and nonlinear warping functions to compare our method with the $k$-means clustering method in \cite{Sang:2010}, which is designed for linear warping functions. The clustering results for the two methods are quite different, as expected.

In Section 3.4, we consider different $q_{1-a}$ values and different clustering index settings. The $q_{1-a}$ values considered are $q_{0. 95}$, $q_{0. 85}$,  $q_{0. 65}$, $q_{0. 55}$ and $q_{0. 45}$. For the clustering index, we try the Dunn index (\cite{Dunn:1974}) under different inter-cluster distances and intra-cluster distances to compare the results with those based on the Silhouette index.
 
 In our simulation experiments,  we use splines to approximation shape curves and warping functions.  All data curves are first approximated using cubic splines, and the  knots are selected using the method proposed by  \cite{Zhou:2001}  with one initial knot at 0.5. Then, we evaluate the approximated curves at 500 equally spaced points in $[0,1]$ to obtain  the apprixomate observed curves and then perform shape curve approximation.  For shape curve approximation, we use cubic splines with 16 equally spaced inner knots, and evaluate shape curves at 500 time points for finding  fitted splines using the method of least squares.  The shape curve approximation is also performed whenever a new shape curve is obtained during curve updating. For the approximation of warping functions, we use quadratic splines with three equally spaced inner knots. We also use quadratic splines with 23 equally spaced inner knots to approximate the inverse of warping functions.

\subsection{Case of warping functions satisfying the boundary condition} \label{section:p1}
We generate three sets of curves $G_1$--$G_3$. For $i \in \{ 1, 2, 3 \}$, the $i$-th set is composed of $N_i$ similar curves, which have a common shape function $f_i$ if properly warped. The shape functions for $G_1$--$G_3$  are  given below:
\[f_1(t)= \sin(2.5 \pi t),\]
\[f_2(t)=(-t^2+\sin(2 \pi t)+0.25)/1.3, \]
and 
\[f_3(t)=\sin(2.5 \pi t^{2.5}), \]
for $t \in [0, 1]$. 
For the warping functions, we consider  functions of the form $t^\alpha$, where $\alpha \in \{ 0.86+0.03(k-1): k=1, \ldots, 10 \}$ so that the boundary condition holds. $N_i$ is either 10 or 20 in this study. For $N_i = 10$, we use warping functions $\psi_1$, $\ldots$, $\psi_{10}$, where $\psi_k(t) = t^{0.86+0.03(k-1)}$ for $t \in [0, 1]$ and $k=1$, $\ldots$, 10. For $N_i = 20$, we use warping functions $\psi_1$, $\ldots$, $\psi_{10}$, $\psi_1$, $\ldots$, $\psi_{10}$. Every curve is generated with equally spaced  time points $t_1=0$, $\ldots$, $t_{100}=1$, and the $j$-th generated curve in the $i$-th group is
\begin{equation} \label{gen data}
y_{ij}(t) = f_i(\psi_j(t)) + \varepsilon_{ij}(t),
\end{equation}
where $\varepsilon_{ij}(t_1)$, $\ldots$, $\varepsilon_{ij}(t_{100})$ are IID $N(0, \sigma^2)$. Here, $\sigma = 0.15$ or $0.45$.

\begin{figure}[ht] \centering 
        \includegraphics[width=0.35 \textwidth]{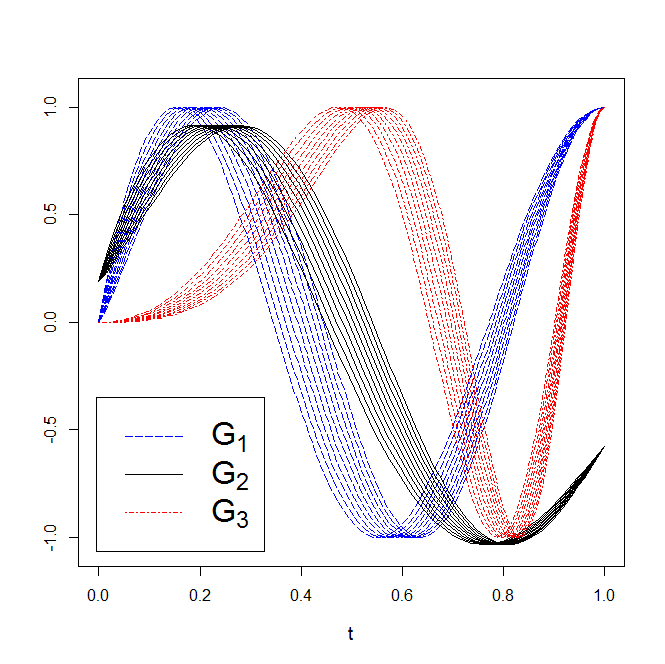} 
     \caption{$G_1$--$G_3$ curves without errors}
     \label{curve30}
     \end{figure}

Note that $f_3(t) = f_1(t^{2.5})$, and thus if $G_1$ and $G_3$ curves are properly warped, they will have the same shape function.
Figure \ref{curve30} shows the three sets of curves without errors. It appears that curves in the same group follow a similar pattern since the time variation within each group is not very large. In contrast, for   $G_1$ and $G_3$ curves , although they have the same shape function when properly warped, the unwarped curves for the two groups show quite different patterns due to large variation in time. 

To investigate the effect of $\lambda_0$ and class size, two values for $\lambda_0$ and seven class sizes are considered, and the adjusted Rand indexes proposed by \cite{Hube:1985} for evaluating clustering results are calculated for the following two cases.
\begin{itemize} 
\item[(a)] There are two clusters $G_1 \cup G_3$ and $G_2$. This clustering case is denoted by $\big((G_1, G_3), G_2, 2\big)$.
\item[(b)] There are three clusters $G_1$, $G_2$, and $G_3$.
 This clustering case is denoted by $\big(G_1, G_2, G_3, 3\big)$.
\end{itemize}
Table \ref{simu1} shows the adjusted Rand index averages of 30 experiments for Cases (a) and (b), and the standard deviations are given in parentheses.  Note that when $\lambda_0=0$, the shape functions for  $G_3$ and $G_1$ curves are perfectly similar according to our similarity measure, and accordingly,  our method usually returns the clustering result $\big((G_1, G_3), G_2, 2\big)$ that matches Case (a).   When $\lambda=0.5$, the shape functions for $G_3$ and $G_1$ curves are less similar since there is a penalty for using warping functions that are different from identity. As a result, our method often returns the clustering result $\big(G_1, G_2, G_3, 3\big)$ matches Case (b). This phenomenon is observed under various combinations of class size and  $\lambda_0$ for each $\sigma$. In addition, the effect of class size is not significant.
\begin{table*}[ht] \renewcommand{\arraystretch}{1.2} \small
\caption{Adjusted Rand index averages for different sizes of groups and $\lambda_0$s} \label{simu1}
\addtolength{\tabcolsep}{-3.5pt}
\begin{center}
    \begin{tabular}{cccccc}
    \hline  \hline
    $\big((G_1, G_3), G_2, 2\big)$ & \multicolumn{2}{c}{$\sigma = 0.15$} && \multicolumn{2}{c}{$\sigma = 0.45$}\\
    \hline
    $(N1,N2,N3)$ & $\lambda_0 = 0.0$ & $\lambda_0 = 0.5$  && $\lambda_0 = 0.0$ & $\lambda_0 = 0.5$\\
    \hline
    $(10,10,10)$ & 1(0) & 0.5538(0) && 0.9802(0.1009) & 0.5513(0.0098) \\
    $(10,10,20)$ & 1(0) & 0.5185(0) && 0.9668(0.1448) & 0.5179(0.0035) \\
    $(10,20,10)$ & 1(0) & 0.7417(0) && 0.9984(0.0089) & 0.7403(0.0076) \\  
    $(20,10,10)$ & 0.9974(0.0141) & 0.5172(0.0073) && 0.9974(0.0141)  & 0.5185(0)  \\
    $(20,20,10)$ & 0.9990(0.0057) & 0.6755(0) && 0.9984(0.0086) & 0.6745(0.0056)  \\
    $(20,10,20)$ & 1(0) & 0.4096(0) && 0.9953(0.0179) & 0.4082(0.0056)  \\
    $(10,20,20)$ & 1(0) & 0.6755(0) && 0.9816(0.0753) & 0.6730(0.0081)   \\
   \hline \hline
   $\big(G_1, G_2, G_3, 3\big)$ & \multicolumn{2}{c}{$\sigma = 0.15$} && \multicolumn{2}{c}{$\sigma = 0.45$}\\
    \hline
    $(N1,N2,N3)$ & $\lambda_0 = 0.0$ & $\lambda_0 = 0.5$ && $\lambda_0 = 0.0$ & $\lambda_0 = 0.5$\\
    \hline
    $(10,10,10)$ & 0.5538(0) & 1(0) && 0.5586(0.0363) & 0.9967(0.0125)\\
    $(10,10,20)$ & 0.5185(0) & 1(0) && 0.5138(0.0182) & 0.9992(0.0046) \\
    $(10,20,10)$ & 0.7417(0) & 1(0) && 0.7399(0.0097) & 0.9985(0.0082) \\  
    $(20,10,10)$ & 0.5175(0.0058) & 0.9982(0.0098) && 0.5175(0.0058) & 1(0) \\
    $(20,20,10)$ & 0.6744(0.0062) & 1(0) && 0.6748(0.0036) & 0.9988(0.0063)  \\
    $(20,10,20)$ & 0.4096(0) & 1(0) && 0.4090(0.0023) & 0.9977(0.0088)  \\
    $(10,20,20)$ & 0.6755(0) & 1(0) && 0.6772(0.0154) & 0.9972(0.0091)  \\
   \hline \hline
\end{tabular}
\end{center}
\end{table*} 
\subsection{Case of  warping functions slightly violating the boundary condition} \label{section:p2}
In this section, we examine the performance of our method when the boundary  condition is violated slightly.  We generate data using shape functions $f_1$, $f_2$, and $f_3$ in Section \ref{section:p1}, but the warping functions are of the following  forms:
\[\psi_a(t)=a_1t+a_2, t \in [0, 1]\]
and
\[\psi_b(t)=(1+b_2-b_1)t^\alpha+b_1, t \in [0, 1],\]
where $a_1, a_2, b_1, b_2$ are generated from uniform distributions $U(0.975, 1.025)$, $U(0, 0.05)$, $U(0, 0.05)$, and $U(-0.05$, $0.05)$, respectively.
The range for $\alpha$ is the same  as in Section \ref{section:p1}. 
Note that $\psi_a$ is a linear function, which is a common choice for warping functions, and $\psi_b$ is a function such that $\psi_b(0)=b_1$ and $\psi_b(1)=1+b_2$. The two types of warping functions do not satisfy the boundary condition if $(a_1, a_2) \neq (1, 0)$ and $(b_1, b_2) \neq (0, 0)$.  For this part of simulation studies, we only consider $(N1, N2, N3)=(10, 10, 10)$ and $\sigma = 0.15$. The clustering results are shown in Table \ref{simu2}.

In Table \ref{simu2}, the adjusted Rand index averages for different combinations of warping functions and $\lambda_0$ values are given for Cases (a) and (b). Due to the violation of the boundary condition,  the clustering performance here is slightly  different from when warping functions satisfy the boundary condition.  However, for the effect of $\lambda_0$, the phenomenon observed in Section \ref{section:p1} is still present here. That is, our method returns results that match the $\big((G_1, G_3), G_2, 2\big)$ case well when $\lambda_0=0$, and it returns results that match the $\big(G_1, G_2, G_3, 3\big)$ case well when $\lambda_0=0.5$.  
\begin{table*}[h]
\caption{Adjusted Rand index averages using warping functions $\psi_a$ and $\psi_b$} \label{simu2}
 \renewcommand{\arraystretch}{1.1} \small
\begin{center}
    \begin{tabular}{ccccc}
    \hline  \hline
    & \multicolumn{2}{c}{$\psi_a$} & \multicolumn{2}{c}{$\psi_b$}\\
    & $\big((G_1, G_3), G_2, 2)$ & $\big(G_1, G_2, G_3, 3\big)$ & $\big((G_1, G_3), G_2, 2)$ & $\big(G_1, G_2, G_3, 3\big)$ \\
    \hline
    $\lambda_0 = 0.0$ & 1(0) & 0.5538(0) & 0.9828(0.0939) & 0.5658(0.0652) \\
    $\lambda_0 = 0.5$ & 0.5538(0) & 1(0) & 0.5538(0) & 1(0) \\
    \hline \hline
    \end{tabular}
\end{center}
\end{table*}

\subsection{Case  of warping assumptions not holding} \label{sec:p3}
In this section, we compare the clustering result of our method with that of the $k$-means clustering method in \cite{Sang:2010}, designed for linear warping functions, under two situations: (1) the warping functions are linear and (2) the warping functions satisfy the boundary condition.  For our method, the warping assumption is violated in Case (1). For the $k$-means method, the warping assumption is violated in Case (2). In what follows, the details of simulation data, clustering results , and some discussions are presented. We use the R package ``fdama" to perform the $k$-means clustering method in \cite{Sang:2010}.

\begin{figure*}[ht] \centering 
\subfigure[]{
    \includegraphics[scale=0.21]{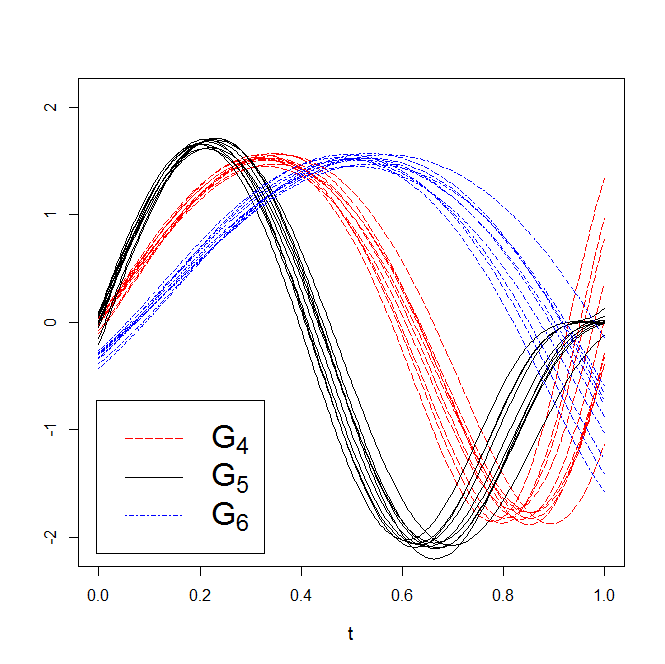} 
}
\subfigure[]{
    \includegraphics[scale=0.21]{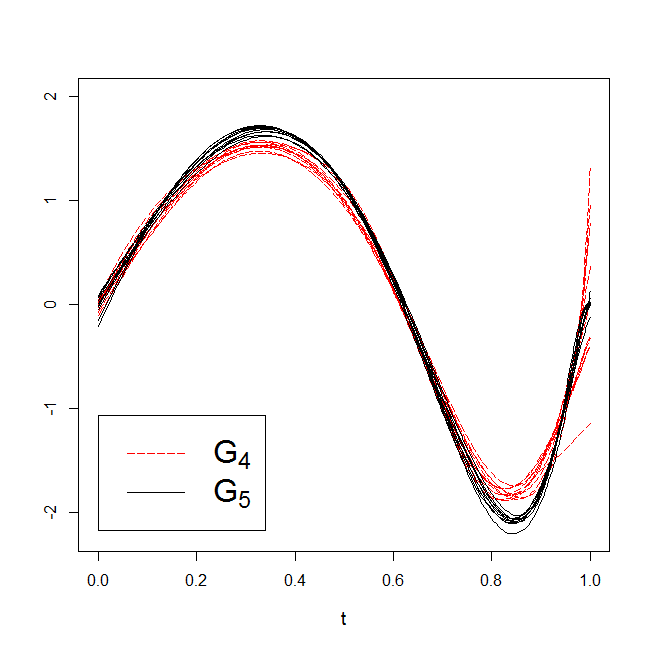} 
}
\subfigure[]{
    \includegraphics[scale=0.21]{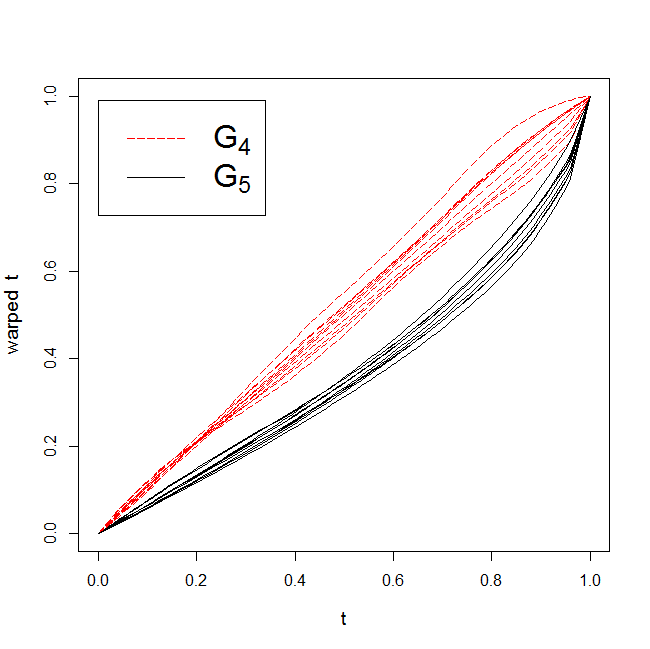} 
}
\caption{Curves and clustering results for $G_4$ -- $G_6$ using our method}
\label{curvekmean}
\end{figure*}

First, we consider Case (1). In this experiment, we generate three groups of simulation curves $G_4$--$G_6$ using random functions $f_4$--$f_6$ given below, which are taken from \cite{Sang:2010} with the modification that a linear function is used to transform the time range from  $[0, 2\pi]$ to $[0,1]$.
\begin{eqnarray*}
 && f_4(t) = (1+\varepsilon_1)\sin \big(\varepsilon_2+(1+\varepsilon_3)2\pi t \big)\\
 && \hspace{0.5cm} +(1+\varepsilon_4)\sin \Bigg(\frac{\big(\varepsilon_2+(1+\varepsilon_3)2\pi t \big)^2}{2\pi}\Bigg), t \in [0, 1],
\end{eqnarray*}
\begin{eqnarray*}
&& f_5(t) = (2+\varepsilon_1)\sin \big(\varepsilon_2+(1+\varepsilon_3)2\pi t \big)\\
 && \hspace{0.5cm} +(-1+\varepsilon_4)\sin  \Bigg(\frac{\big(\varepsilon_2+(1+\varepsilon_3)2\pi t \big)^2}{2\pi}\Bigg), t \in [0, 1],
\end{eqnarray*}
and
\begin{eqnarray*}
&& f_6(t) =  (1+\varepsilon_1)\sin \bigg(\frac{-1}{3}+\varepsilon_2+\big(\frac{3}{4}+\varepsilon_3\big)2\pi t \bigg) \\
& &  \hspace{0.2cm} +(1+\varepsilon_4)\sin \bigg(  \frac{1}{2\pi}\Big( \displaystyle 
-\frac{1}{3}+ \varepsilon_2+\big(\frac{3}{4}+\varepsilon_3\big)2\pi t \Big)^2 \bigg), t \in [0,1].
\end{eqnarray*}
For each of $G_4$--$G_6$, 10 curves are generated using  $f_4$--$f_6$ respectively, but the random errors in $G_4$ are the same as those in $G_5$ and $G_6$. 
The curves in $G_4$ and $G_6$ can be synchronized using linear warping functions. The graph of the data curves are shown in  Figure \ref{curvekmean}(a). 
For the $k$-means clustering method, the curves can be clustered into two groups: $G_4 \cup G_6$ and $G_5$ when the initial centers are specified properly. This result is expected since the curves in $G_4$ and $G_6$ can be synchronized by linear warping functions.  

For our clustering method with $\lambda_0 = 0.0$, we obtain two groups: $G_4 \cup G_5$  and $G_6$. This clustering result is quite different from that of the $k$-means method, because the curves in $G_4$ and $G_5$ have similar patterns when they are aligned using nonlinear warping functions satisfying the boundary condition. Figure \ref{curvekmean}(c) shows the warping functions when the curves in $G_4$ and $G_5$ are aligned to a reference curve and 
Figure \ref{curvekmean} (b) shows the aligned curves. The curves in $G_6$ cannot be aligned well to the curves in $G_4$ or $G_5$ using warping functions satisfying the boundary condition, and as such, they are not clustered into one group under  our method.
 
In the second experiment, we consider Case (2). We use shape functions 
\[ g_1(t) = \sin(2 \pi t^2), t \in [0, 1], \]
\[ g_2(t) = \cos(2 \pi t^2), t \in [0, 1] \] 
and warping functions $\{\varphi_1, \ldots, \varphi_4\} =\{ t^{0.78}, t^{0.89}, t^{1.11}, t^{1.22}\}$ to generate two groups of simulation curves  $G_7$ and $G_8$. Each group is composed of four curves with the same shape function, generated according to  (\ref{gen data}) but without errors. 

In the second experiment, our clustering method with $\lambda_0 = 0.0$, gives the clustering result of two groups $G_7$ and $G_8$. We also apply the $k$-means clustering method in \cite{Sang:2010} with the initial number of groups ranging from one to five. 

\begin{figure}[ht] \centering 
        \includegraphics[width=0.5 \textwidth]{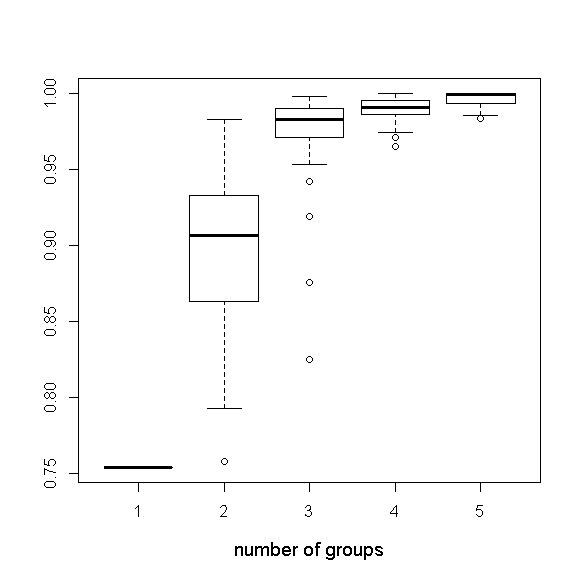} 
     \caption{Clustering results for $G_7$--$G_8$ using $k$-means clustering in \cite{Sang:2010}}
     \label{boxplot}
     \end{figure}

For a given number of clusters $k$, we use every possible combination of $k$ curves among all data curves as initial cluster centers and obtain the average similarity measures between curves and their cluster centers for the corresponding clustering result.  Figure \ref{boxplot} shows the box plot of averages of similarity measures for each $k \in \{ 1, \ldots, 5 \}$. 
This figure shows that the median of averages of similarity measures increases as $k$ increases, and $k$ needs to be at least 3 for the averages of similarity measures to be less sensitive to the choice of initial cluster centers. The result for the $k$-means method is quite different from that for our method since all the curves in $G_7$ (or $G_8$) cannot be aligned well to each other using linear warping functions.

\begin{figure}[h] \centering 
        \includegraphics[width=0.45 \textwidth]{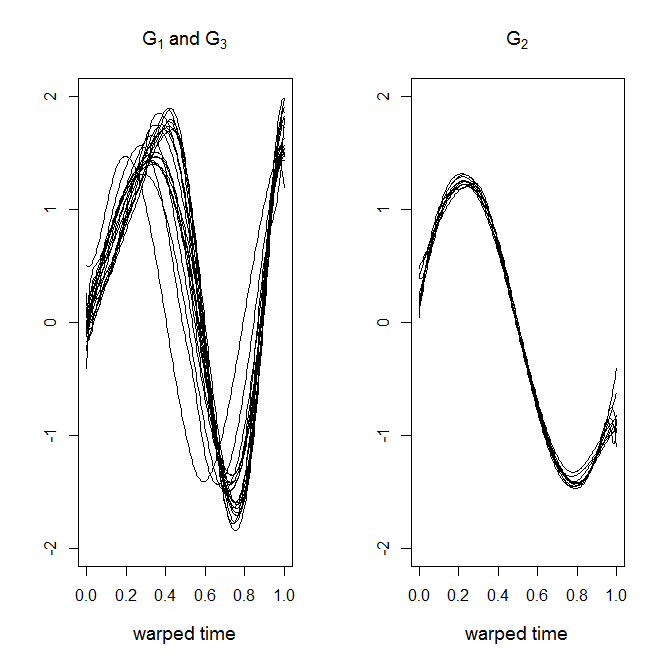} 
     \caption{Warped curves after the first-stage alignment}
     \label{tangplot}
     \end{figure}

In addition to the above experiments, we also tried the two-stage approach proposed in \cite{Tang:2009}. In the first stage, the cluster-specific warping functions were estimated using our similarity measure.  In the second stage, we used the function ``kmeans.fd'' in R package ``fda.usc'' to perform $k$-means clustering to the aligned curves. The clustering results were not as good as ours in terms of the adjusted Rand index averages for the $\lambda_0=0$ case. In addition, for the case  $\lambda_0=0$ and $\sigma=0.15$, the perfect clustering result is $(G_1 \cup G_3, G_2)$ and our method gives the perfect result in all of  the 30 trials. However, the two-stage approach gives the perfect result in only 13 of  the 30 trials.  Figure \ref{tangplot} shows the warped curves after the first-stage alignment using the two-stage approach.  Note that the warped curves in $G_1 \cup G_3$  show larger variation in time as compared to  the warped curves in $G_2$. As a result, $k$-means clustering usually assigns the curves in $G_2$ into the same cluster, but fails to assign the curves in $G_1 \cup G_3$ into the same cluster.
\subsection{Effect of $q_{1-a}$ and clustering index} \label{sec:p4}
In this section, we investigate the influence of  $q_{1-a}$ and clustering index on the proposed clustering method. The simulation data here are the same as those in the case $(N1, N2, N3) = (10, 10, 10)$ in Section \ref{section:p1}. 

First, we apply the proposed method to the simulation data under different $q_{1-a}$ values. Table \ref{different_c} shows the average Adjusted Rand index corresponding to five  $q_{1-a}$  values under $\sigma = 0.15$ and  $0.45$.  We find that the performance of the proposed method is still satisfactory  when $q_{1-a} \in \{  q_{45}, q_{55}, q_{65}, q_{85} \}$.  When $q_{1-a} = q_{95}$, the proposed method sometimes gives a large number of clusters. This is probably due to the fact that the curves cannot be combined in few iterations and  the limit for the number of iterations is set to 10. 
\begin{table*}[ht]
\caption{Adjusted Rand index averages for different $q_{1-a}$ values and $\lambda_0$s} \label{different_c}
 \renewcommand{\arraystretch}{1.2} \small
\addtolength{\tabcolsep}{-3.5pt}
\begin{center}
    \begin{tabular}{cccccc}
    \hline  \hline
    $\big((G_1, G_3), G_2, 2\big)$ & \multicolumn{2}{c}{$\sigma = 0.15$} && \multicolumn{2}{c}{$\sigma = 0.45$}\\
    \hline
    $q_{1-a}$ & $\lambda_0 = 0.0$ & $\lambda_0 = 0.5$  && $\lambda_0 = 0.0$ & $\lambda_0 = 0.5$\\
    \hline
    $q_{95}$ & 0.9728(0.1492) & 0.5538(0) && 0.6606(0.4551) & 0.5076(0.1112) \\
    $q_{85}$ & 1(0) & 0.5538(0) && 0.9986(0.0076) & 0.5538(0) \\
    $q_{65}$ & 1(0) & 0.5538(0) && 0.9986(0.0076) & 0.5538(0) \\  
    $q_{55}$ & 1(0) & 0.5538(0) && 0.9986(0.0076) & 0.5538(0)  \\
    $q_{45}$ & 1(0) & 0.5538(0) && 0.9931(0.0307) & 0.5538(0)  \\
   \hline \hline
   $\big(G_1, G_2, G_3, 3\big)$ & \multicolumn{2}{c}{$\sigma = 0.15$} && \multicolumn{2}{c}{$\sigma = 0.45$}\\
    \hline
    $q_{1-a}$ & $\lambda_0 = 0.0$ & $\lambda_0 = 0.5$ && $\lambda_0 = 0.0$ & $\lambda_0 = 0.5$\\
    \hline
    $q_{95}$ & 0.5493(0.0251) & 1(0) && 0.4136(0.2027) & 0.9303(0.1702) \\
    $q_{85}$ & 0.5538(0) & 1(0) && 0.5523(0.0087) & 1(0) \\
    $q_{65}$ & 0.5538(0) & 1(0) && 0.5523(0.0087) & 1(0) \\  
    $q_{55}$ & 0.5538(0) & 1(0) && 0.5523(0.0087) & 1(0)  \\
    $q_{45}$ & 0.5538(0) & 1(0) && 0.5519(0.0088) & 1(0)  \\
   \hline \hline
\end{tabular}
\end{center}
\end{table*} 

Next, we apply the proposed method  using the Dunn index to compare the clustering results with those based on the Silhouette index. The Dunn index based on groups $G_1$, $\ldots$, $G_k$ is defined by
\[  \frac{\min \limits_{1 \leq i < j \leq k} d_{inter}(G_i, G_j)}{\max \limits_{1 \leq i \leq k} d_{intra}(G_i)},\]
where $d_{inter}$ and $d_{intra}$ are inter-cluster distance and intra-cluster distance respectively.  In this simulation study, we use three inter-cluster distances (I1)--(I3) and two intra-cluster distances (J1)--(J2) when evaluating the Dunn index.  The definitions of these distances can be found in \cite{Sole:2013} and stated below. In the following descriptions, $d(a,b)$ denotes the distance between two elements $a$ and $b$ and $|A|$ denotes the number of elements in group $A$.
\begin{itemize}
\item $d_{inter}(G_i, G_j)$
\begin{itemize}
\item[(I1):] $\min \limits_{x_1 \in G_i, x_2 \in G_j} d(x_1, x_2)$
\item[(I2):] $\max \limits_{x_1 \in G_i, x_2 \in G_j} d(x_1, x_2)$
\item[(I3):] $\frac{1}{|G_i||G_j|}\sum d(x_1, x_2)$, where $x_1 \in G_i$ and $x_2 \in G_j$
\end{itemize}
\item $d_{intra}(G_i)$
\begin{itemize}
\item[(J1):] $\max \limits_{x_1, x_2 \in G_i, x_1 \neq x_2} d(x_1, x_2)$
\item[(J2):] $\frac{1}{|G_i|(|G_i|-1)}\sum d(x_1, x_2)$, where $x_1, x_2 \in G_i$ and  $x_1 \neq x_2$
\end{itemize}
\end{itemize}
 In Table \ref{dunn}, we show the adjusted Rand index averages for the proposed method using Dunn index with the inter-cluster distances and intra-cluster distances mentioned above. Only the case $(\sigma, q_{1-a})=(0.15,  q_{75})$ is considered. We find that the  Adjusted Rand index averages are similar for the six cases $(I1, J1)$ -- $(I3, J2)$, and the clustering results are  very similar to the results based on Silhouette index. 
\begin{table*}[ht]
\caption{Adjusted Rand index averages based on Dunn index} \label{dunn}
 \renewcommand{\arraystretch}{1.2} \small
\addtolength{\tabcolsep}{-3.5pt}
\begin{center}
    \begin{tabular}{ccccccc}
    \hline  \hline
   $\big((G_1, G_3), G_2, 2\big)$ & $(I1, J1)$ & $(I1, J2)$ & $(I2, J1)$  & $(I2, J2)$ & $(I3,J1)$ & $(I3, J2)$\\
    \hline
    $\lambda_0 = 0.0$ & 1(0) & 1(0) & 0.9971(0.0159) & 1(0) & 0.9971(0.0159) & 1(0) \\
    $\lambda_0 = 0.5$ & 0.5538(0) & 0.5538(0) & 0.5538(0) & 0.5538(0) & 0.5538(0) & 0.5538(0)\\
   \hline \hline
   $\big(G_1, G_2, G_3, 3\big)$ & $(I1, J1)$ & $(I1, J2)$ & $(I1, J3)$  & $(I2, J1)$ & $(I2,J2)$ & $(I2, J3)$\\
    \hline
    $\lambda_0 = 0.0$ & 0.5538(0) & 0.5538(0) & 0.5535(0.0019) & 0.5538(0) & 0.5535(0.0019) & 0.5538(0) \\
    $\lambda_0 = 0.5$ & 1(0) & 1(0) & 1(0) & 1(0) & 1(0) & 1(0) \\
   \hline \hline
\end{tabular}
\end{center}
\end{table*}

%% file: data_analysis.tex
In this section, we apply our method to two data sets using $q_{1-a}=q_{75}$.
\subsection{Berkeley growth study data} \label{sec:4.1}
\begin{figure*}[h] \centering 
\subfigure[]{
    \includegraphics[scale=0.23]{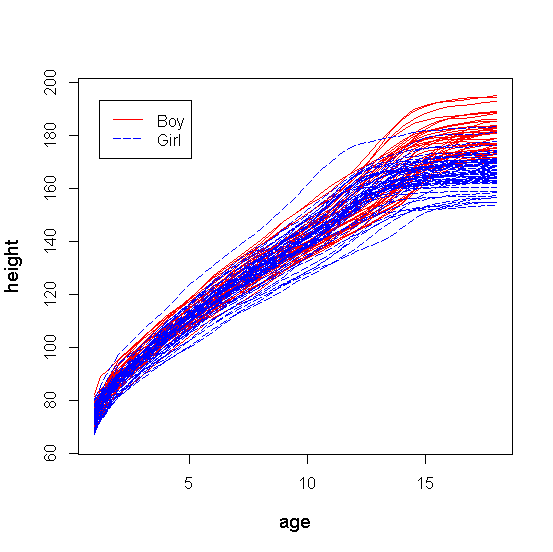} 
}
\subfigure[]{
    \includegraphics[scale=0.23]{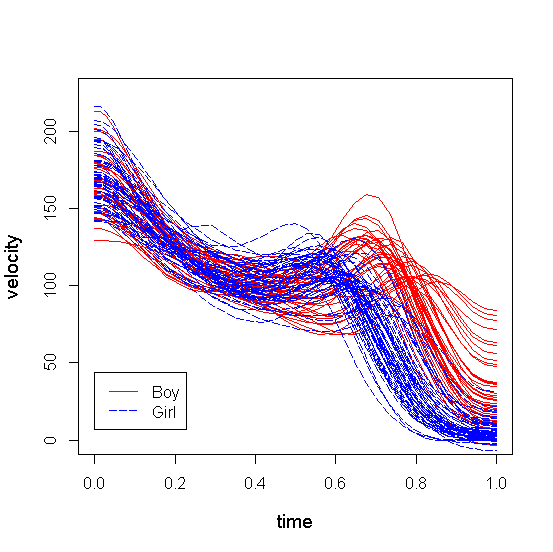} 
}
\caption{Height and velocity curves for the Berkeley growth study}
\label{growth_obs}
\end{figure*}

\begin{figure*}[h] \centering
\subfigure[Clustering result ($\lambda_0=0$)]{
   \label{fig:subfig1}
   \includegraphics[scale=0.22]{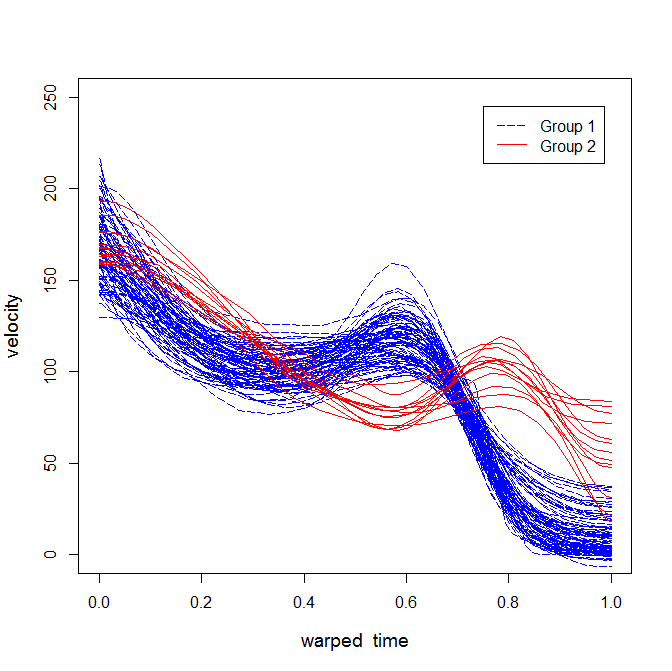}
}
\subfigure[Warping functions ($\lambda_0=0$)]{
   \label{fig:subfig2}
   \includegraphics[scale=0.22]{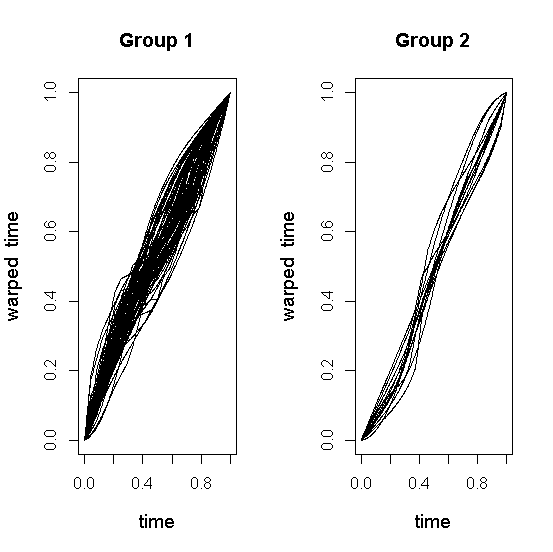}
}
\subfigure[Clustering result ($\lambda_0=0.5$)]{
   \label{fig:subfig3}
   \includegraphics[scale=0.22]{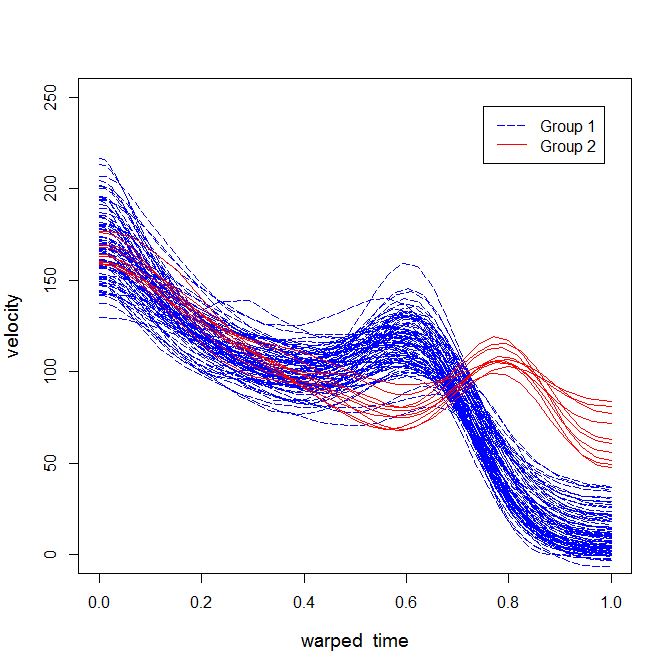}
}
\subfigure[Warping functions ($\lambda_0=0.5$)]{
   \label{fig:subfig4}
   \includegraphics[scale=0.22]{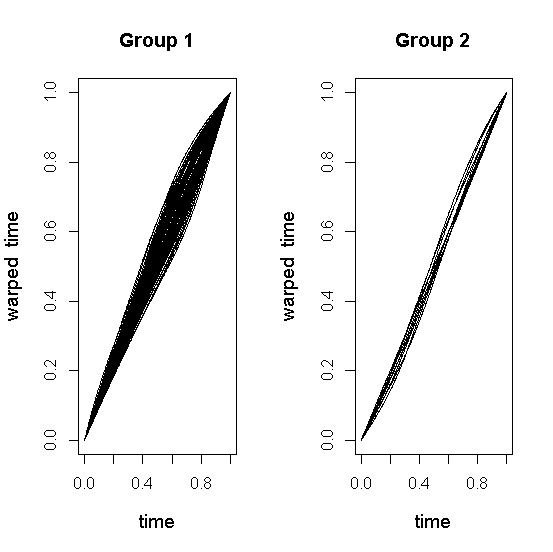}
}
 \caption{Clustering results and warping functions for growth study data}
 \label{growth_res}
\end{figure*}

The Berkeley growth study data set (\cite{Tudd:1954}) consists of the heights of 39 boys and 54 girls at 31 time points from 
when they were a year old to when they were 18 years old. Figure \ref{growth_obs}(a) shows these height curves, and Figure (b) shows the corresponding growth velocity curves that are obtained using the smoothing technique in Section 4.2 in \cite{Rams:1997}. In this analysis, we apply the proposed clustering method to the growth velocity curves instead of the original height curves. The height curves are increasing functions of age, and thus they inevitably form similar curves  with $\lambda_0=0$. For convenience, we apply a linear function to transform the age range $[1, 18]$ into $[0, 1]$.  
Figure \ref{growth_res} shows the clustering results of these velocity curves from our clustering method with $\lambda_0 = 0.0$ and $0.5$. In both cases, these velocity curves are classified into two groups. We show the aligned velocity curves in Figures \ref{growth_res}(a) and (c). 

In Figures \ref{growth_res}(a) and (c), for the curves in Group 2, the growth velocities at the end are greater than those in Group 1. It seems that the boys/girls whose growth curves are in Group 2 continued to grow higher at great speed at age 18 and their growth velocity curves show a different pattern since the growth processes were not yet complete. With $\lambda_0=0.5$, we found that Group 1 contains growth curves of boys and girls, but Group 2 contains growth curves of only boys. 
Figures \ref{growth_res} (b) and (d) show the warping functions for $\lambda_0=0$ and 0.5. The warping functions for $\lambda_0 = 0.5$ are smoother than those for $\lambda_0 = 0$, as expected.

\subsection{Baby Finder data} \label{sec:4.2}
A Baby Finder is an electronic device comprising a receiver and a transmitter, where the transmitter can send signals to the receiver continuously. The Baby Finder data set contains signal loss data for a Baby Finder from eight trials, provided by the third author of this paper. 
For each trial, the transmitter and the receiver are put together first and then are taken away along two paths respectively. If the moving path pairs for two trials are the same, we expect the corresponding signal loss curves to be similar.
\begin{figure*}[h] \centering 
\subfigure[]{
    \includegraphics[scale=0.21]{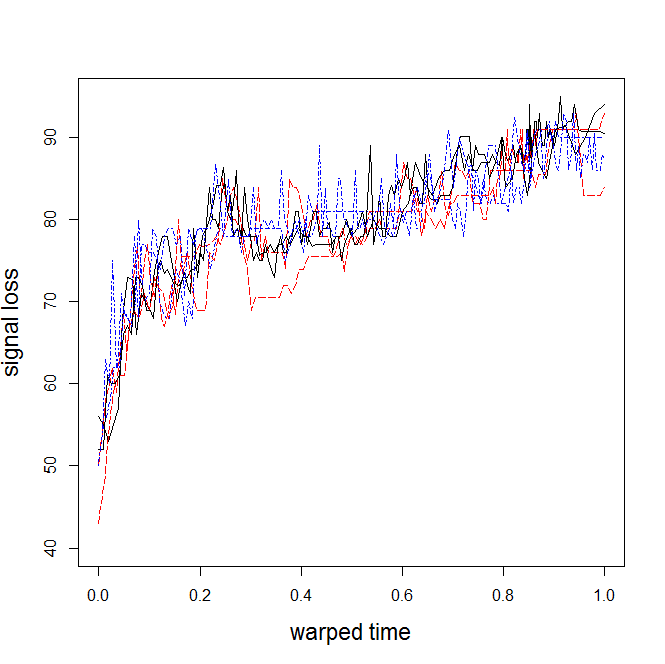} 
}
\subfigure[]{
    \includegraphics[scale=0.21]{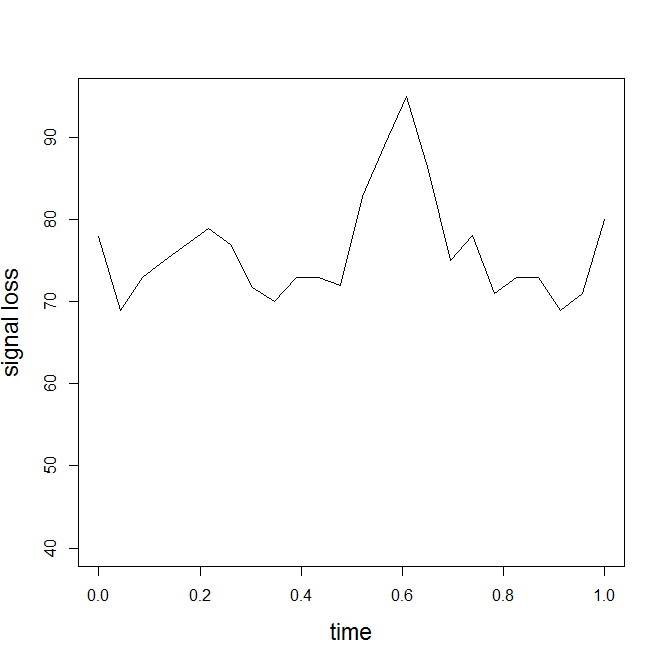} 
}
\subfigure[]{
    \includegraphics[scale=0.21]{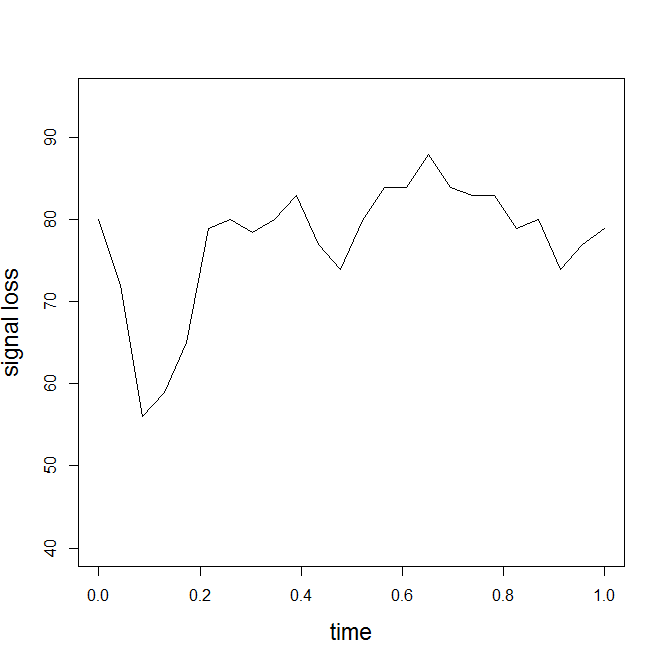} 
}
\caption{Clustering results for Baby Finder data}
\label{babyfinder}
\end{figure*}  
Applying the proposed clustering method to those signal loss curves, the eight curves are clustered into three groups. The curves in the three groups are shown in Figures \ref{babyfinder}(a)--(c), respectively. The left figure shows the aligned six curves for the first group,  all of which increase as time increases. The rest two curves have different patterns corresponding to different moving paths, and thus  the two are clustered into two singleton groups.  For each of the  six curves in the first group, in addition to signal loss measurements, we have information on the travel distances of the receiver at different time points. We find that  for the six trials, the signal losses at the same distance are about the same, and the maximum travel distances are the same, so we suspect that  the receiver and the transmitter were taken away from the  same pair of paths respectively for the six trials. The clustering results support our guess.

%% file: conclusion.tex
Based on our simulation results, the proposed method works well when warping functions satisfy the boundary condition. Regarding the implementation, one needs to choose  $S_c$, the penalty parameter $\lambda_0$ and the clustering index. Below are our comments and suggestions when $S_c$ is determined using $q_{1-a}$.
\begin{itemize}
\item When $q_{1-a}$ is small, it takes very few iterations for the curves to be combined, but curves that are not very similar may be combined and may be put into the same cluster due to the small combination threshold values.  As $q_{1-a}$ increases, it can be ensured that only similar curves are combined but it takes more iterations for the curves to be combined.  Based on our simulation experiments, the clustering results remain stable when $q_{1-a}$
  is between $q_{45}$ and $q_{85}$. If $q_{1-a}$ is too large, the curves may not be combined in the limited number of iterations and one may have to increase the limit for the number of iterations, and computation time can be longer. Our suggestion is to choose a large $q_{1-a}$ (as large as possible, but not large enough to make the number of iterations reaches its limit). 

\item In most of our simulation experiments, we use $q_{1-a} = q_{75}$, and the computation takes a lot of time.  For the case $(N1, N2, N3)=(10, 10,10)$ and $\lambda_0=0$ in Table 1,  the median compuation times (three trials)  for $\sigma=0.45$ and $\sigma=0.15$ are  22142 seconds and 7574 seconds respectively using a machine with Intel CPU i7-4790K. For the case where $(N1, N2, N3)=(10, 10,20)$, $\lambda_0=0$ ,  $\sigma=0.15$, the median compuation times (three trials) is 11133 seconds. If we use $q_{1-a}= q_{55}$, for the case where $(N1, N2, N3)=(10, 10,10)$, $\lambda_0=0$ ,  $\sigma=0.15$, the median compuation time (three trials) becomes 3970 seconds, which is much less than that of the case with  $q_{1-a}= q_{75}$. 

\item The choice of $\lambda_0$ depends on to what extent time variation is considered as a clustering factor.  Consider the case where one curve $y_1$ can be perfectly aligned to another curve $y_2$ using a warping function $\psi$,  but $\psi$ is very different from the identity function. In such case, if one would like $y_1$ and $y_2$ to be assigned into the same cluster (time variation is not important), then $\lambda_0$ should be set to 0. Otherwise, a nonzero $\lambda_0$ should be used. Using a large $\lambda_0$ means that time variation  is considered as an important factor in clustering.

\item Since the results based on the Dunn index are very similar to those based on the Silhouette index,  the effect of clustering index does not seem to be significant. If one would like to choose a clustering index, it is recommended to use a clustering index that does not involve cluster centers (such as the  Silhouette coefficient or the Dunn index) since the proposed method does not compute cluster centers in the clustering process.

\end{itemize}

%% file: proof.tex
In the section, we give the proof of Theorem \ref{thm1}, and the proofs of two facts: Facts 1 and 2, which are used in the proof of Theorem \ref{thm1}.  We will state and prove Facts 1 and 2 first. 

We first state Fact \ref{fact1}. Let $V = L^2[0,1]$. Below are the assumptions, statement and proof of Fact \ref{fact1}.
\begin{itemize}
\item[(A1)] Suppose that $h_1$, $\ldots$, $h_k \in V$ and $g=\sum_{j=2}^k \theta_j h_j$, where $\theta_j$s are constants in $[0,1]$ such that $\sum_{j=2}^k \theta_j=1$.
\item[(A2)]  Suppose  that for $j =2$, $\ldots$, $k$, $\langle \cdot, \cdot \rangle_j$ is a positive semi definite symmetric bilinear form from $V \times V$ to $R$. 
\item[(A3)] For $j =2$, $\ldots$, $k$, let $\|  \cdot \|_j$  be the semi-norm on $V$ defined by $\| f \|_j = \sqrt{\langle f, f \rangle_j}$ for $f \in V$.
\item[(A4)] Suppose that $\| h_j \|_j= 1$ for $j=1$, $\ldots$, $k$.
\item[(A5)] Suppose that $c_1$, $\ldots$, $c_k$ are positive constants.
\end{itemize}
\begin{fact} \label{fact1}
Suppose that (A1)--(A5) hold. For $\lambda >0$, let 
\[
 T(\lambda) = \sum_{j=2}^k \frac{c_j \langle \lambda h_1 + g, h_j \rangle_j }{ \| \lambda h_1  + g \|_j }\]
 and
 \[T(\infty) = \lim_{\lambda \goes \infty} T(\lambda) = \sum_{j=2}^k \frac{c_j \langle h_1, h_j \rangle_j }{ \| h_1 \|_j } .
\]
Let  $\displaystyle A_j = \frac{c_j \langle h_1, h_j \rangle_j }{ \| h_1 \|_j } $,  
  $\displaystyle B_j = \frac{c_j \langle g,  h_j \rangle_j }{ \| h_1 \|_j } $, 
 $\displaystyle  D_j = \frac{ 2\langle h_1, g \rangle_j }{ \| h_1 \|_j^ 2 } $,
 $\displaystyle  E_j = \frac{\| g \|_j}{ \| h_1 \|_j}$,  
 $\displaystyle \alpha_j = B_j - \frac{1}{2} A_j D_j $ and
\[
 \beta_j = \frac{1}{2} \left( A_jE_j^2 + B_jD_j + |B_j| E_j \right) + \frac{3}{\sqrt{2}} (|A_j|+1)  (|D_j|+E_j)^2.
\]
If $\sum_{j=2}^k \alpha_j >0$ and 
\begin{equation} \label{eq:lambda}
 \lambda \geq \max \left( \frac{ \sum_{j=2}^k \beta_j}{ \sum_{j=2}^k \alpha_j }, \max_{2 \leq j \leq k } \max(E_j, |B_j|) \right),
\end{equation}
then $T(\lambda) \geq T(\infty)$.
\end{fact}
{\bf Proof of Fact \ref{fact1}.}
Note that 
\[
  T(\lambda) = \sum_{j=2}^k \frac{\lambda A_j + B_j }{ \sqrt{ \lambda^2 + \lambda D_j + E_j^2} } \mbox{ and } T(\infty) = \sum_{j=2}^k A_j.
\]
 Let $\displaystyle U_j =  \frac{\lambda A_j + B_j }{ \sqrt{ \lambda^2 + \lambda D_j + E_j^2} } - A_j$, then $T(\lambda) -T(\infty) = \sum_{j=2}^k U_j$.

To find a lower bound for the expression
\[
  U_j = \frac{ \displaystyle A_j + \frac{B_j}{\lambda} }{ \displaystyle \sqrt{ 1 + \frac{D_j}{\lambda} + \frac{E_j^2}{\lambda^2} } } - A_j,
\]
we consider the Taylor expansion of $1/\sqrt{1+x}$ at $x=0$, which gives
\[
  \frac{1}{\sqrt{1+x}} = 1- \frac{1}{2} x+ \frac{1}{2} \cdot \frac{3}{4} (1+c)^{-5/2} x^2,
\]
where $c$ is between $0$ and $x$.  Apply the result from Taylor expansion with  $\ds x=\xo$ and we have
\begin{eqnarray}
 U_j & = & A_j \left( 1- \frac{1}{2} \left( \xo \right) + \frac{3\tilde{c} }{8} \left( \xo\right)^2 \right) \nn \\
 && + \frac{B_j}{\lambda} \left( 1 - \frac{1}{2} \left( \xo \right) + \frac{3\tilde{c} }{8}  \left( \xo \right)^2 \right) \nn \\
 &&- A_j, \label{eq:uj}
\end{eqnarray}
where $\tilde{c} =  (1+c)^{-5/2}$. Suppose that (\ref{eq:lambda}) holds, then $\lambda \geq E_j$, using the fact that $|D_j| \leq 2E_j$, we have 
\[
 -\frac{1}{2} \leq  \xo \leq 3,
\]
which implies that $1+c \geq 0.5$ and $0 < \tilde{c} \leq 2^{5/2}$. It then follows from (\ref{eq:lambda}) and (\ref{eq:uj}) that
\begin{eqnarray*}
U_j & \geq & \frac{1}{\lambda} \left( B_j - \frac{1}{2} A_j D_j \right) \\
 && - \frac{1}{2} \frac{A_j E_j^2}{ \lambda^2} - \frac{3\tilde{c}}{8} |A_j| \left(\frac{|D_j|+E_j}{ \lambda } \right)^2 \\
 && -\frac{1}{2\lambda^2} (B_j D_j + |B_j|E_j) -  \frac{3\tilde{c}}{8} \frac{|B_j|}{\lambda} \frac{ (|D_j|+E_j)^2}{\lambda^2} \\
 & \geq & \frac{1}{\lambda} \left( B_j - \frac{1}{2} A_j D_j \right) \\
 && -\frac{1}{\lambda^2} \left( \frac{1}{2}( A_jE_j^2 + B_j D_j + |B_j|E_j) \right) \\
 && -\frac{1}{\lambda^2}\left( \frac{3(2^{5/2})}{8} (|A_j|+1)(|D_j|+E_j)^2\right) \\
 & = &  \frac{\alpha_j}{\lambda} -\frac{\beta_j}{\lambda^2}.
\end{eqnarray*}
Thus for  $\lambda \geq \sum_{j=2}^k \beta_j/\sum_{j=2}^k \alpha_j$, $\sum_{j=2}^k U_j \geq 0$. Therefore,  $T(\lambda)-T(\infty) = \sum_{j=2}^k U_j \geq 0$ when (\ref{eq:lambda}) holds. The proof of Fact \ref{fact1}  is complete.

Next, we give the assumptions, statement and proof of Fact \ref{fact2}.
\begin{itemize}
\item[(A6)] Suppose that $\langle \cdot, \cdot \rangle$ is a positive semi definite symmetric bilinear form from $V \times V$ to $R$. 
\item[(A7)] Let $\|  \cdot \|$  be the semi-norm on $V$ defined by $\| f \| = \sqrt{\langle f, f \rangle}$ for $f \in V$.
\item[(A8)] Suppose that $\| h_j \|= 1$ for $j=1$, $\ldots$, $k$.
\end{itemize} 
\begin{fact} \label{fact2}
Suppose that (A1) and (A6)--(A8) holds. Let $s = \sum_{j=2}^k h_j$. For $\lambda >0$, let 
\[
 T(\lambda) = \sum_{j=2}^k \frac{ \langle \lambda h_1 + g, h_j \rangle}{ \| \lambda h_1  + g \| } = \frac{ \langle \lambda h_1 + g, s \rangle}{ \| \lambda h_1  + g \| } 
\]
and
\[
 T(\infty) = \lim_{\lambda \goes \infty} T(\lambda) = \sum_{j=2}^k \frac{\langle h_1, h_j \rangle }{ \| h_1 \| }  = \sum_{j=2}^k \langle h_1, h_j \rangle = \langle h_1, s  \rangle.
\] 
Let \[r(g) = g - \langle g, h_1 \rangle h_1\] and
\[r(s) = s - \langle s, h_1 \rangle h_1.\]
If $\langle h_j, h_1 \rangle \geq 0$ for every $j \geq 2$, $\langle r(g), r(s) \rangle >0$, $\lambda >0$ and
\begin{equation} \label{eq:lambda2}
 \lambda \geq  \frac{\| r(g) \|^2 \langle s, h_1 \rangle^2 - \langle r(g), r(s) \rangle^2}{ 2\langle s, h_1 \rangle \langle r(g), r(s) \rangle} - \langle g, h_1 \rangle,
\end{equation}
then $T(\lambda) \geq T(\infty)$.
\end{fact}
{\bf Proof of Fact \ref{fact2}.}
Let $\beta = \lambda + \langle g, h_1 \rangle$, then
\begin{eqnarray*}
T(\lambda) - T(\infty) & = & \frac{\squareo}{\rooto} -  \langle s, h_1 \rangle \\
&=& \frac{\ds \left( \frac{\squareo}{\rooto} \right)^2 - \langle s, h_1 \rangle^2}{ \frac{\squareo}{\rooto} + \langle s, h_1 \rangle }.
\end{eqnarray*}
Under the conditions that $\lambda >0$, $\langle h_j, h_1 \rangle \geq 0$ for every $j \geq 2$ and $\langle r(g), r(s) \rangle >0$, we have $\langle s, h_1 \rangle \geq \| h_1 \|^2 =1$ and \[ \frac{\squareo}{\rooto} + \langle s, h_1 \rangle >0, \] 
 so
\begin{eqnarray*}
&& T(\lambda) \geq T(\infty) \\
&& \Leftrightarrow  (\squareo)^2 - (\beta^2 + \| r(g) \|^2) \langle s, h_1 \rangle^2 \geq 0 \\
& & \Leftrightarrow  2\beta  \langle s, h_1 \rangle   \langle r(g), r(s) \rangle \geq   -  \langle r(g), r(s) \rangle ^2 + \| r(g)\|^2  \langle s, h_1 \rangle ^2  \\
 && \Leftrightarrow (\ref{eq:lambda2}) \mbox{ holds. }
\end{eqnarray*}
 The proof of Fact \ref{fact2}  is complete. 

Next, we provide the proof of Theorem \ref{thm1} as follows. \\
{\bf Proof of Theorem \ref{thm1}}. To establish 
\begin{equation} \label{eq:main}
 \sum_{j=2}^k \rho (f_1, f_j)  \leq \sum_{j=2}^k \rho( \lambda f_1 + g_0, f_j) = \sum_{j=2}^k \rho ( f^*_1, f_j),
\end{equation} 
we will prove
\begin{equation} \label{eq:main1}
  \sum_{j=2}^k r(f_1, f_j \circ \psi_j)  \leq  \sum_{j=2}^k r ( \lambda f_1 + g_0, f_j \circ \psi_j) 
\end{equation}
and
\begin{equation} \label{eq:main2}
 \sum_{j=2}^k r ( f_1 \circ \psi_j^{-1}, f_j)  \leq \sum_{j=2}^k r ( ( \lambda f_1 + g_0) \circ \psi_j^{-1}, f_j).
\end{equation}
Then 
\small
\begin{eqnarray*}
&& \sum_{j=2}^k \rho (f_1, f_j) =   \sum_{j=2}^k \rho (f_1, f_j|\psi_j) \\
  &&=  \sum_{j=2}^k \frac{1}{2} \left( r(f_1, f_j\circ \psi_j) - \lambda_0 \int_0^1 \left( \frac{d}{dt} \psi_j(t) - 1 \right)^2 dt \right) \\
 && + \sum_{j=2}^k \frac{1}{2} \left( r (f_1\circ \psi_j^{-1}, f_j) - \lambda_0  \int_0^1 \left( \frac{d}{dt} \psi_j^{-1}(t) - 1 \right)^2 dt \right) \\
& & \stackrel{(\ref{eq:main1}), (\ref{eq:main2})}{\leq}  \sum_{j=2}^k \rho(\lambda f_1 + g_0, f_j|\psi_j) \\
 && \leq \sum_{j=2}^k \rho( \lambda f_1 + g_0, f_j)
\end{eqnarray*}
\normalsize
and (\ref{eq:main}) holds.

To prove (\ref{eq:main1}), note that  (\ref{eq:main1}) follows from Fact \ref{fact2} with $h_1=f_1$, $g=g_0$ and $\ds h_j = \frac{f_j \circ \psi_j}{\| f_j \circ \psi_j \|}$ for $j \geq 2$.  The required conditions are implied by (C1)--(C3) and (C5).

To prove (\ref{eq:main2}), we will apply Fact \ref{fact1}. For $f\in V$,
\[
  r(f \circ \psi_j^{-1}, f_j) = \frac{\inn{f}{f_j\circ \psi_j}_j}{ \| f \|_j}.
\]
Apply Fact \ref{fact1} with $h_1=f_1$, $c_j = \| f_j \circ \psi_j \|$  and $h_j = f_j \circ \psi_j/c_j$ for $j \geq 2$. 
Then (\ref{eq:main2}) holds if   $\sum_{j=2}^k \alpha_j >0$ and (\ref{eq:lambda}) holds. Note that $\sum_{j=2}^k \alpha_j >0$ is equivalent to the condition (C4) and (\ref{eq:lambda})  is given in (C6). Therefore, (\ref{eq:main2}) holds.

%% file: paper.bbl
\begin{thebibliography}{17}
\providecommand{\natexlab}[1]{#1}
\providecommand{\url}[1]{\texttt{#1}}
\expandafter\ifx\csname urlstyle\endcsname\relax
  \providecommand{\doi}[1]{doi: #1}\else
  \providecommand{\doi}{doi: \begingroup \urlstyle{rm}\Url}\fi

\bibitem[Chen and Shiu(2007)]{Chen:2007}
T.-L. Chen and S.-Y. Shiu.
\newblock A new clustering algorithm based on self-updating process.
\newblock In \emph{Proceedings of the American Statistical Association,
  Statistical Computing Section [CD-ROM]}, Salt Lake City, Utah. (2007).

\bibitem[Dunn(1974)]{Dunn:1974}
J.~C. Dunn.
\newblock Well separated clusters and optimal fuzzy partitions.
\newblock \emph{Journal of Cybernetics}, 4:\penalty0 95--104 (1974).

\bibitem[Fukunaga and Hostetler(1975)]{Fuku:1975}
K.~Fukunaga and L.~D. Hostetler.
\newblock The estimation of the gradient of a density function, with
  applications in pattern recognition.
\newblock \emph{IEEE Transactions on Information Theory}, 21:\penalty0 32--40
  (1975).

\bibitem[Gervini and Gasser(2004)]{Gerv:2004}
D.~Gervini and T.~Gasser.
\newblock Self-modelling warping functions.
\newblock \emph{Journal of the Royal Statistical Society, Series B: Statistical
  Methodology}, 66:\penalty0 959--971 (2004).

\bibitem[Hubert and Arabie(1985)]{Hube:1985}
L.~Hubert and P.~Arabie.
\newblock Comparing partitions.
\newblock \emph{Journal of Classification}, 2:\penalty0 193--218 (1985).

\bibitem[Jacques and Preda(2014)]{Jacq:2014}
J.~Jacques and C.~Preda.
\newblock Functional data clustering: a survey.
\newblock \emph{Advances in Data Analysis and Classification}, 8:\penalty0 231--255 (2014).

\bibitem[James(2007)]{Jame:2007}
G.~M. James.
\newblock Curve alignment by moments.
\newblock \emph{The Annals of Applied Statistics}, 1:\penalty0 480--501 (2007).

\bibitem[Kneip and Gasser(1992)]{Knei:1992}
A.~Kneip and T.~Gasser.
\newblock Statistical tools to analyze data representing a sample of curves.
\newblock \emph{The Annals of Statistics}, 20:\penalty0 1266--1305 (1992).

\bibitem[Kneip et~al.(2000)Kneip, Li, MacGibbon, and Ramsay]{Knei:2000}
A.~Kneip, X.~Li, K.~B. MacGibbon, and J.~O. Ramsay.
\newblock Curve registration by local regression.
\newblock \emph{The Canadian Journal of Statistics/La Revue Canadienne de
  Statistique}, 28:\penalty0 19--29 (2000).

\bibitem[Liu and Yang(2009)]{Liu:2009}
X.~Liu and M.~C. Yang.
\newblock Simultaneous curve registration and clustering for functional data.
\newblock \emph{Computational Statistics \& Data Analysis}, 53:\penalty0
  1361--1376 (2009).

\bibitem[Ramsay and Li(1998)]{Rams:1998}
J.~O. Ramsay and X.~Li.
\newblock Curve registration.
\newblock \emph{Journal of the Royal Statistical Society, Series B: Statistical
  Methodology}, 60:\penalty0 351--363 (1998).

\bibitem[Ramsay and Silverman(1997)]{Rams:1997}
J.~O. Ramsay and B.~W. Silverman.
\newblock \emph{Functional Data Analysis}.
\newblock Springer-Verlag Inc (1997).
\newblock ISBN 0-387-94956-9.

\bibitem[Rousseeuw(1987)]{Rous:1987}
P.~J. Rousseeuw.
\newblock Silhouettes: A graphical aid to the interpretation and validation of
  cluster analysis.
\newblock \emph{Journal of Computational and Applied Mathematics}, 20:\penalty0
  53 -- 65 (1987).

\bibitem[Sangalli et~al.(2010)Sangalli, Secchi, Vantini, and
  Vitelli]{Sang:2010}
L.~M. Sangalli, P.~Secchi, S.~Vantini, and V.~Vitelli.
\newblock $k$-mean alignment for curve clustering.
\newblock \emph{Computational Statistics \& Data Analysis}, 54:\penalty0
  1219--1233 (2010).

\bibitem[Shiu and Chen(2012)]{Shiu:2012}
S.-Y. Shiu and T.-L. Chen.
\newblock Clustering by self-updating process.
\newblock In \emph{arxiv:1201.1979} (2012).

\bibitem[Silverman(1995)]{Silv:1995}
B.~W. Silverman.
\newblock Incorporating parametric effects into functional principal components
  analysis.
\newblock \emph{Journal of the Royal Statistical Society, Series B:
  Methodological}, 57:\penalty0 673--689 (1995).

\bibitem[Soler et~al.(2013)Soler, Tenc{\'e},  Gaubert, and Buche]{Sole:2013}
 J.~Soler, F.~Tenc{\'e}, L.~Gaubert, and C.~Buche.
\newblock Data clustering and similarity.
\newblock In {\em Proceedings of the Twenty-Sixth International Florida
  Artificial Intelligence Research Society Conference}, pages 492--495, St.
  Pete Beach, Florida. (2013).


\bibitem[Tang and M{\"u}ller(2009)]{Tang:2009}
R.~Tang and H.-G. M{\"u}ller.
\newblock Time-synchronized clustering of gene expression trajectories.
\newblock \emph{Biostatistics}, 10:\penalty0 32--45 (2009).

\bibitem[Telesca and Inoue(2008)]{Tele:2008}
D.~Telesca and L.~Y.~T. Inoue.
\newblock Bayesian hierarchical curve registration.
\newblock \emph{Journal of the American Statistical Association}, 103:\penalty0
  328--339 (2008).

\bibitem[Tuddenham and Snyder(1954)]{Tudd:1954}
R.~D. Tuddenham and M.~M. Snyder.
\newblock Physical growth of california boys and girls from birth to eighteen
  years.
\newblock In \emph{University of Califormia Publications in Child Development},
  volume~1, pages 183--364. University of California Press (1954).

\bibitem[Zhou and Shen(2001)]{Zhou:2001}
S.~Zhou and X.~Shen.
\newblock Spatially adaptive regression splines and accurate knot selection schemes.
\newblock \emph{Journal of the American Statistical Association}, 96:\penalty0
  247--259 (2001).

\end{thebibliography}
